# Graphical abstract

## 2D-Double Transition Metal MXenes for Spintronics Applications: Surface Functionalization Induced Ferromagnetic Half-Metallic Complexes

Kripa Dristi Dihingia[a,b], Swagata Saikia[a], N. Yedukondalu[a], Supriya Saha[a,b,*], G. Narahari Sastry[a,b]

[a]Advanced Computation and Data Sciences Division, Council of Scientific and Industrial Research-North East Institute of Science and Technology (CSIR-NEIST), Jorhat, 785006, Assam, India.
[b]Academy of Scientific and Innovative Research (AcSIR), Ghaziabad, 201002, Uttar Pradesh, India.

[*]E-mail: supriya.saha@neist.res.in

## Table of content entry

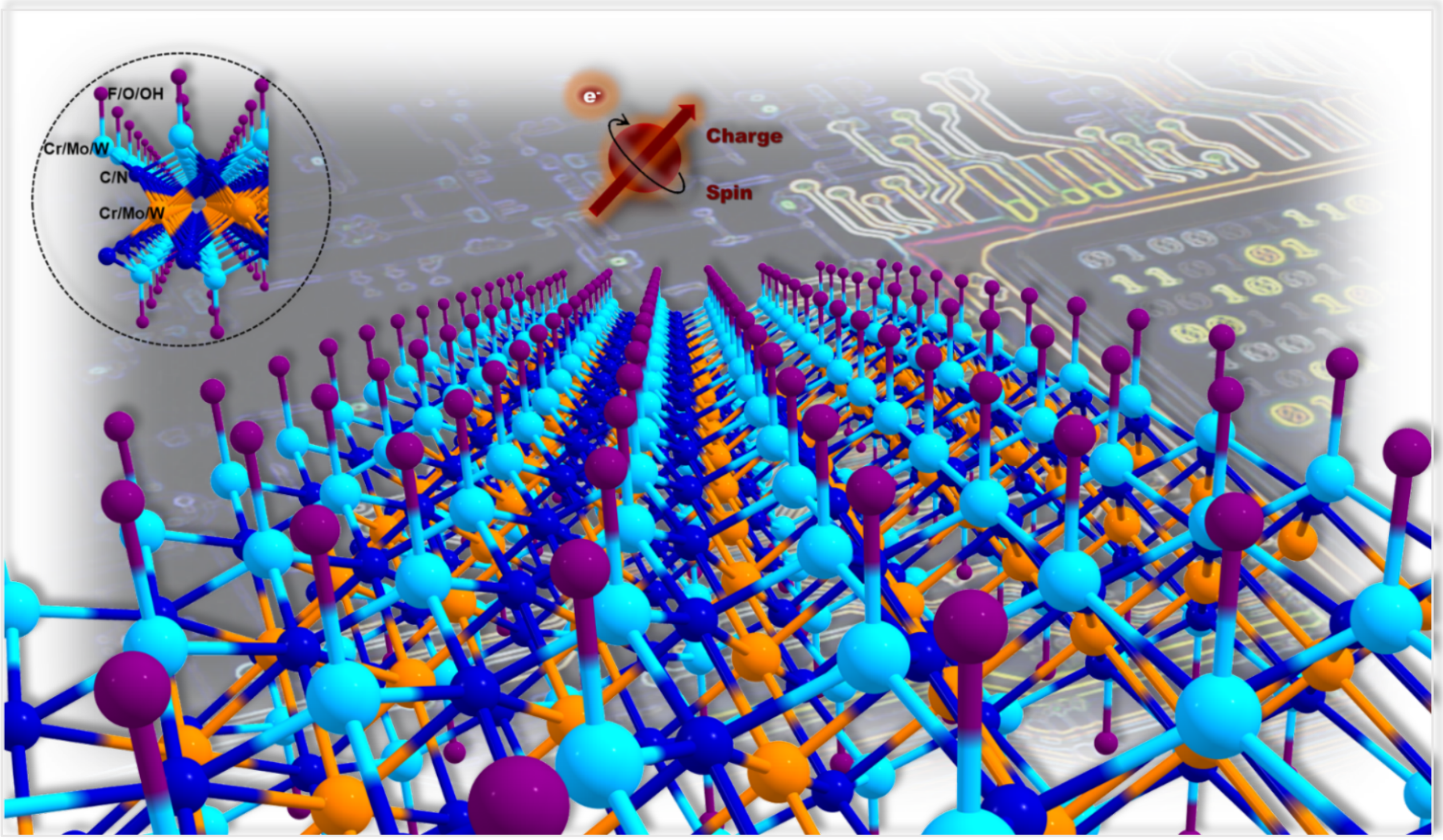

# 2D-Double Transition Metal MXenes for Spintronics Applications: Surface Functionalization Induced Ferromagnetic Half-Metallic Complexes

Kripa Dristi Dihingia[a,b], Swagata Saikia[a], N. Yedukondalu[a], Supriya Saha[a,b,*], G. Narahari Sastry[a,b]

[a]Advanced Computation and Data Sciences Division, Council of Scientific and Industrial Research-North East Institute of Science and Technology (CSIR-NEIST), Jorhat, 785006, Assam, India.
[b]Academy of Scientific and Innovative Research (AcSIR), Ghaziabad, 201002, Uttar Pradesh, India.

[*]E-mail: supriya.saha@neist.res.in

## Abstract

MXenes are rapidly emerging two-dimensional (2D) materials with thickness, composition, and functionalization-dependent outstanding properties having applications in diverse fields. To disclose nano-spintronic applications of 2D-double transition metal (DTM) carbide and nitride-based pristine and surface-functionalized MXenes ($M'_2M''X_2T_x$, M′ and M′′ = Cr, Mo, W; X = C/N; T = –F/–OH/=O), a systematic investigation has been performed on structural stability, magnetic properties and electronic structure using spin-polarized first-principles calculations. 36 stables functionalized MXenes were screened from 144 explored DTM based MXenes. The explored materials exhibit striking properties, having wide range of magnetic ground states, from non-magnetic to ferromagnetic, and then to antiferromagnetic, accompanied by metallic to half-metallic or gapless half-metallic properties, depending on transition metal(s) and terminating group. Mo and W-based MXenes are found to be nonmagnetic and metallic, whereas Cr-Mo and Cr-W-based MXenes are magnetic with varying metallic behavior. $W_2CrN_2O_2$ and $Mo_2CrN_2O_2$ systems are found to be ferromagnetic half-metallic 2D materials with a direct band gap of 1.35 eV and 0.77 eV respectively, in the minority spin channel. The comprehensive study on DTM MXenes, provide intrinsic half-metallic properties along with robust ferromagnetism, opens up a class of promising new 2D materials with tunable magnetic and electronic properties for potential device applications in nano-spintronics and electronics.

## 1. Introduction

Since electrons have two degrees of freedom, *i.e.*, charge and spin, integration of semiconducting microelectronics (based on the charge of electrons) and magnetic storage devices (relay on the spin of electrons) into a single chip significantly advances information technology.[1] As the conventional charge-based electronics reached a hard limit (Ex. 7 nm Fin Field Effect Transistor and further reduction of transistor node size to 1 nm will end the progress of electronic devices) to further improve it, therefore, it is indispensable to look for an alternative. Spintronics might be an alternative next-generation technology, which can use the spin of electrons in addition to the charge as information carriers to speed up the data processing with low energy consumption.[2–4] However, the discovery of functional spintronic materials (half metals, spin gapless semiconductors (SCs), bipolar magnetic SCs, etc.) with curie temperature above 300 K is a challenge.[5,6]

Since the successful exfoliation of graphene from graphite,[7] the discovery of advanced 2D-materials (h-BN, Phosphene, transition metal dichalcogenides (TMDs), etc.) has been fueled up. Eventually it turned out to be an era of infinitely vigorous exploration of functional 2D materials.[8–10] Due to the quantum confinement effect and large surface-to-volume ratio, 2D materials possess outstanding properties in contrast to their three-dimensional (3D) counterparts and have wide range of applications.[11,12] However, in the field of nano-spintronics, applications of low-dimensional materials are limited owing to lack of intrinsic magnetism.[13–17]

The vast majority of the experimentally produced 2D crystals are nonmagnetic. For the application of 2D materials in the field of electronics and spintronics, opening and tuning of the bandgap and magnetic moment in controllable way is essential. This includes different techniques like size quantization, variations of compositions, chemical functionalization with different functional group, by inducing strain and electric field and so on.[17–19]

After the room temperature exfoliation of $Ti_3C_2$ from $Ti_3AlC_2$ in 2011 by Naguib and co-workers,[20] the 2D MXenes (transition metal carbides or nitrides or carbonitrides) are emerging as a formidable and extensive family of materials with fascinating features, which can promise new scientific phenomenon and wide range of technological horizons.[21–26] The MXenes are derived by etching the "A" element from MAX phases, which have the chemical formula of $M_{n+1}AX_n$ (n = 1-3) (where, M = early transition metal; A = traditionally group 13-16 elements, and X = C or/and N).[20,27–29] MXenes remarkably opened up potential possible applications in spintronic devices,[2–4,17] owing to their numerous possible compositions with various structural frameworks, and combined with unique electronic, and magnetic properties.[18,30,31] MXenes are unique 2D materials because of 1) Chemically active MXene surface can be functionalized with diverse chemical groups (–F, –OH, =O, –H, –Cl, –S, –Se, –Te, etc.), which offers an avenue for surface state engineering to discover new MXenes for diverse technological applications ranging from nanoelectronics to energy storage and conversion,[32–35] 2) Thickness of the monolayer is controllable based on the number of transition metal atoms, along with scaling from monolayers to multilayers analogous to other 2D materials,[36] and 3) Manipulations of the transition metal layers.[37]

In the present work, DTM-based carbide and nitride MXenes were considered, possess a general formula $M'_2M''X_2T_x$, where M′ and M′′ = Cr, Mo, W; X = C or N and T = –F, –OH, or =O, respectively. DTM MXenes possess interesting physical and chemical properties than their counterpart mono transition metal based MXenes and their properties can be tuned by controlling the compositions.[31,38] The research and development on DTM MXenes are in their early stages and many compositions yet to be explored computationally and novel MXenes could be synthesized. Besides, recent research mainly focused on DTM carbide MXenes where exploration on nitride DTM MXenes are limited and carbonitride DTM MXenes have yet to be explored.[39–42] Depending upon the atomic lattice structures and the compositions of MXene, these materials can be differentiated into ordered mono- and double-M, and solid-solution M elements.[43] The modelled systems in this work exist in ordered phases with one transition metal layer lies between the layers of another transition metal. Depending on the ordering pattern of transition metal atoms, MAX phases can be distinguished into out of plane (o-MAX) and in plane (i-MAX) ordered systems.[42,43] In this work, investigated pristine and functionalized $Cr_2MoX_2$, $Cr_2WX_2$, $Mo_2CrX_2$, $Mo_2WX_2$, $W_2CrX_2$, and $W_2MoX_2$ (X = C or N) MXenes, can be exfoliated as out of plane ordered MXene (o-MXene) from o-MAX phases. It has been already established that the ordered DTM MXenes are structurally stable.[42] Majority of the MXenes, composed with early transition metals, have less number of electrons in the valence orbital. Hence, the number of unpaired electron spins are limited and show metallic and/or non-magnetic (NM) behavior. MXenes composed with Group-VI transition metals, possess more number of valence electrons in d-orbitals than early transition metals and considered as possible 2D magnetic materials.[31] In 2013, Khazaei and co-workers theoretically predicted $Cr_2CT_x$ ($T_x$ : –F, –OH) and $Cr_2NT_x$ ($T_x$ : –F, =O, –OH) can show magnetic behavior with surface functionalization, although this study couldn't clarify the type of magnetism, either FM or AFM for systems under consideration.[44] Si *et al.* proposed a new promising material $Cr_2C$, which possesses ferromagnetism and intrinsic half-metallicity with tunable magnetic (FM to AFM transition) properties from the calculated electronic properties for possible applications towards spin-electronic devices and this is the first reported half-metallic ferromagnet MXene.[45] Afterwards, several, $Ti_2C$ and $Ti_2N$,[46] $Cr_3C_2$,[47] $Mn_2CF_2$,[48] $Mn_2NT_2$ (T= O, F, OH),[49] $Ti_2NO_2$,[49] $TiMn_2C_2F_2$,[50] $Cr_2NT_2$ (T = O, F, OH),[51,52] $Cr_2CuC_2$,[6] $Nd_2NT_2$ (T = O, S, Cl, OH, F and Br)[53] MXenes includes in this family.

With this motivation, we shed more light on investigating structural stability, magnetic properties and electronic structure of DTM (Cr, Mo, and W) carbide and nitride based MXenes with surface functionalization using various chemical groups (–F, –OH, or =O) for their possible potential applications in nanoscale spintronics and electronics. The detailed workflow of the explored complex systems for screening the stable functionalized MXenes, and their possible potential applications based on their magnetic properties and electronic structures, is depicted in Fig. 1.

## 2. Computational Details

In this work, a spin-polarized first-principles calculations were performed in the framework of density functional theory (DFT) to explore the electronic structure and magnetic properties of DTM carbide and nitride MXenes ($M'_2M''X_2$; $M'/M''$ = Cr, Mo, W; X = C/N) considering surface termination (T) with –F, –OH, or =O groups. Materials consist of transition metals found to be tricky for first-principles calculations due to the strongly correlated and localized nature of '*d*' electrons. However, a recent comprehensive study on MXenes using four different functionals (PBE, PBE+U, HSE06, SCAN) revealed that PBE might be sufficient for structural stability and HSE06 for the electronic structure of MXenes.[54] Therefore, in this work, Perdew-Burke-Ernzerhof (PBE)[55] generalized gradient approximation (GGA) was used as an exchange-correlation functional to treat electron-electron interactions. Electron-ion interactions are captured with the projector augmented wave (PAW) approach as implemented in the Vienna ab initio simulation package (VASP).[56,57] The conjugate gradient algorithm is used for full structural optimization where the force criterion for energy convergence was set to $10^{-4}$ eV/Å. The kinetic plane-wave cut-off energy was set to 400 eV. During structural optimization, Γ-centred 12×12×1 k-points grids were adopted. A vacuum space of more than 15 Å was considered along the out-of-plane direction to avoid the spurious interactions. To account the van der Waals (vdW) interactions, the DFT-D3 method with Becke-Johnson damping function has been adopted[58]. For crystal structure visualization and analysis, VESTA,[59] and Chemcraft[60] software packages were used. To verify the dynamic stability of the investigated systems, lattice dynamical calculations were performed by employing the finite displacement method with 5×5×1 supercell and k-grid of 2×2×1 and post-processed using the PHONOPY code[61,62]. For phonon calculations, the atomic configurations were further optimized considering the convergence criterion of the electronic self-consistent cycle below $10^{-6}$ eV and forces smaller than 0.01 eV/Å. Generally, the standard DFT functionals (PBE), underestimate the band gap by 30-40% due to self-interaction error (SIE). To overcome the SIE problem, Heyd-Scuseria-Ernzerhof (HSE06)[63] type hybrid functional has been used to investigate the electronic structure such as band structure, total and projected density of states (DOS). For all the stable relaxed systems, Γ-centered 25×25×1 k-mesh was considered to compute the total and projected DOS. To calculate different magnetic configurations and spin density distribution, 2×2×1 supercell was considered where the Brillouin zone is integrated with a 6×6×1 k-point mesh.

## 3. Results and Discussion

### 3.1. Structural properties

The modelled structures for pristine and passivated MXenes are presented in Fig. 2. As shown in Fig. (2a, 2b), $M'_2M''X_2$ forms a hexagonal lattice with $P\bar{3}m1$ symmetry and comprised of quintuple layers of M′-X-M′′-X-M′ (Fig. 2b). The structure shows that C or N layers are sandwiched between two different transition metal layers (M′ and M′′) and located at the centre of an octahedral cage formed by M′ and M′′. In this DTM MXene system, one layer of transition metal (M′′) is sandwiched between two layers of C or N and other transition metal layers (M′) are located on the surface sites, and they are chemically reactive.

As a first step, full structural relaxation has been adopted on lattice constants and fractional co-ordinates of the constructed pristine $M'_2M''X_2$ systems. The optimized structural parameters are provided in Table 1. The lattice parameter *a* for DTM MXenes slightly differ from single transition metal (STM) based MXenes (Table 1) where calculated lattice parameters for STM

MXenes are consistent with previously reported data.[64] It has been found that the lattice constant for nitride-based MXenes are smaller than carbide-based MXenes (except for $Mo_2WX_2$). This indicates, the structure of nitride based MXenes are more compact than carbide based MXenes. This carbide or nitride based structural compactness order follow the similar pattern with other transition metal (e.g. Ti) MXenes.[65] From structural relaxation it has been observed (Table 1) that interlayer distances between surface metal atoms and X atoms (M′-X) are smaller in comparison with the distances between middle layer transition metals and X atoms (M′′-X) for most of the systems. Similar kind structural relaxation also observed previously by Chen *et al.*[66] for $Cr_2TiC_2T_2$ and $Mo_2TiC_2T_2$ (T = –F, =O, –OH) MXenes where $d_{(Cr/Mo\text{-}C)} < d_{(Ti\text{-}C)}$. The atomic distances between M′-X and M′′-X layers differ depending on the substituted X atoms (C/N) and transition metal atoms from their counterpart mono-transition metal system. Considering the dispersion corrections, the interatomic distances was also explored and the results were presented in Table 1. From the results it has been observed that with and without dispersion corrections for the studied pristine systems the interatomic distances are very close. This indicates that dispersion correction has not much effect on the structural parameters for the studied systems.

To understand the structural stability of DTM MXenes in comparison with STM-based systems, cohesive energies ($E_{coh}$) were calculated. This is a very important parameter to determine the phase stability, as it measures the binding strength of atoms together in the crystal. The cohesive energies of explored pristine systems was calculated using the following equation [64,67,68]:

$$E_{coh} = E_{tot_M'_nM''X_n} - nE_{M'_atom} - E_{M''_atom} - nE_{X_atom} \quad (1)$$

where $E_{tot_M'_nM''X_n}$ is the total energy of $M'_nM''X_n$ system. $E_{M'_atom}$, and $E_{M''_atom}$ are the atomic energy for M′, and M′′, respectively. $E_{X_atom}$ is the energy of X atom. *n* indicating the total number of M′ and C/N atom in the unit cell (for this work, n = 2). To normalize the cohesive energy for different systems, cohesive energy per atom was calculated by dividing $E_{coh}$ with total number of individual constituent atoms (m) in the system, where $\bar{E}_{coh} = \frac{E_{coh}}{m}$ and tabulated in Table 1. The obtained negative normalized cohesive energies ($\bar{E}_{coh}$) for all the modelled systems as shown in Table 1, indicates the structural stability of the considered MXenes. For the formation of $M'_2M''X_2$ system from $M_3X_2$ system, upon replacement of M atom with M′′ by moving down a group in the periodic table, $\bar{E}_{coh}$ decreases, indicates DTM-MXenes are thermodynamically more favorable than STM-MXene counterpart. However, upon replacement of M atom with M′′ by moving up a group, DTM-MXene get destabilise a bit than STM based MXenes. In $Cr_3X_2$ system, upon replacement of middle Cr layer by Mo or W atom, it stabilises the system more than single Cr-based MXenes. Further, in $Mo_3X_2$ and $W_3X_2$ system, on replacement of middle Mo/W-layer with Cr-layer, stability of the system decreases slightly than pure Mo/W-based MXenes, although the $\bar{E}_{coh}$ for all explored DTM MXenes are highly negative. This confirm their formation and phase stability.

As discussed previously, MXenes are constructed by exfoliation of element "A" from MAX phases. The formed pristine MXenes are chemically terminated and hence they are highly reactive to form stable structures upon surface functionalization with various functional groups. To synthesis MXenes from MAX phase, mainly utilized acids are hydrofluoric acid or a mixed acid, solution of hydrochloric acid and lithium fluoride. In this synthesis process, surface atoms mainly terminated by –F, –OH, or =O groups.[28,69] On the MXene surface, there are two distinct hollow sites (Fig. 2a), $H_M$ and $H_X$. $H_M$ sites correspond to face centred cubic (FCC) sites with no C or N atom under M′ atoms and $H_X$ sites represent the hexagonal close packed (HCP) sites placed on hollow-top of C or N atom in $M'_2M''X_2$ system. For the termination of pristine MXenes with functional groups, four model structures were constructed as shown in Fig. 2c (Model I-IV). This enhances the chemical space and we modelled 144 configurations for surface terminated MXenes. On MXene surface, the position of termination for different functional groups plays a major role on their dynamic stability and electronic structure. Therefore, it is very important to determine the energetically most stable configuration for each functional group. Full surface coverage was considered, two terminating groups per unit cell; as fully functionalized MXenes are more stable than partially passivated MXenes confirmed by previous studies.[44] A systematic investigation was performed for all the modelled structures and screened the energetically stable configurations for each functional group. For this, static energy calculations were performed considering full structural relaxation of the systems within PBE method and conducted relative stability energy calculations to detect the most stable configurations for each functionalized MXene. The relative stability energy values for each terminating group are summarized in Table S1-S3, considering four surface termination sites on MXene surface. Based on energetically most stable structure for each terminating group, 36 stable complex structures were tabulated along with preferable adsorption site in Table 2.

It has been observed that for $Cr_2Mo$, $Cr_2W$ and $Mo_2Cr$ based carbide and nitride MXenes, irrespective of surface terminating groups, Model IV are relatively less favourable over the remaining three models (Model I-III). This result is consistent with the single transition metal (Group-VI) based MXenes.[52,70] For $Cr_2M''X_2$ (where M′′ = Mo/W; X = C/N) systems, FCC is the preferred site for surface terminations with =O and –OH groups while –F terminating group favour the FCC-HCP mixing mode. $Cr_2MoC_2F_2$ system formed stable structure with HCP-site. The metal-top site is not preferable for these systems. In $W_2M''X_2$ (where M′′ = Cr/Mo; X = C/N) systems, –F and –OH terminations prefer metal top site (Model IV) whereas =O passivation prefers HCP-site. In case of $Mo_2WX_2$ (X =C/N)-based MXenes, for –F and =O functionalization, most stable structures generated with HCP-site while –OH functionalization prefers metal-top site. For $Mo_2CrX_2$ (X =C/N) systems, =O and –OH terminations favor HCP-site while F-group prefers FCC-HCP mixing site, except $Mo_2CrN_2(OH)_2$ system (FCC-site). The preferable adsorption sites for –F, –OH and =O groups on Cr, Mo, W based DTM-MXenes are analogous with previous studies based on DTM-MXenes.[26,44] Previously, Khazaei *et al.*[44] explored structural stability of $M_2X$ (M = Cr, Zr, Ta, Sc, Ti, V, and Nb; X = C or N) system with –OH, –F, and =O surface terminations and observed that most stable $M_2XT_2$ systems are generated upon termination at FCC or HCP position and these systems do not prefer metal-top site. She *et al.* worked on $M_2XT_2$ [where, M = Zr, Hf, Sc, V, Ti, Ta, Nb, Cr, Mo, and W; X = C or N; and $T_x$ = OH, O, H and $H_2O$] systems and reported that for W, Cr, and Mo-based MXenes, terminated groups favour the HCP-site, while for V, Sc, Ti, Hf, and Zr-based MXenes, prefer FCC-site for surface terminations.[26] We have also explored the stable structures and relative energies incorporating the dispersion correction for all the passivated composite systems and presented in Table S1-S6. It has been observed that for passivated systems, the most stable surface adsorption sites are remain same after inclusion of dispersion correction and the relative energies for adsorption of surface terminating groups at different surface sites are also very close. This indicates that dispersion correction is not affecting the results much from PBE calculations.

To ascertain the structural stability, and formation of $M'_2M''X_2T_x$ systems are thermodynamically feasible or not, adsorption energies were calculated using well-established equation for MXenes in the literature[67]:

$$E_{ads} = \frac{E_{tot_M'_nM''X_nT_x} - E_{tot_M'_nM''X_n} - \frac{x}{2}E_{tot_T_2}}{2} \quad (2)$$

where, $E_{tot_M'_nM''X_nT_x}$ stands for the total energy of the functionalized MXenes while $E_{tot_M'_nM''X_n}$ is the total energy for the pristine systems. *n* indicating the total number of M′ and C/N atom in the unit cell. $x$ stands for the number of passivated groups (–F/–OH/=O) per unit cell. For this current work, the value of n and x is 2. $E_{tot_T_2}$ is the total energy of $F_2$, ($O_2$ + $H_2$), and $O_2$. In each unit cell there are two active surface sites, top and bottom surfaces for terminations. Thus, 2 in the denominator of eqn (2), was considered to execute adsorption energy per surface for the terminations with functionalized groups. Based on the adsorption energies, 36 stable structures were screened among 144 passivated complex structures and provided the adsorption energies for stable surface passivated MXenes in Table 2. The large negative adsorption energy values are indicating the thermodynamic stability of modelled pristine MXenes upon functionalization. It is signifying that the functional groups are attaching strongly with surface metal atoms.

During comparison of stability order of the modelled MXenes in terms of passivating groups, it is found that =O functionalized MXenes are the most stable systems. –F functionalized systems are the 2$^{nd}$ most stable and –OH group passivated MXenes are found to be least stable systems except $Cr_2Mo$/W-based carbide and nitride systems. For $Cr_2Mo$/W-based carbide MXenes, –F passivated systems are most stable followed by –OH functionalized systems and =O functionalized systems, respectively. Besides, for $Cr_2Mo$/W-based nitride-MXenes the stability order for adsorption of functional groups is: –F > =O > –OH. For $Mo_2W$ and $Mo_2Cr$ based carbide and nitride MXenes, functionalized with =O and –F are energetically competitive. Analogy of adsorption energies ($E_{ads}$) for $M'_2M''X_2T_2$ systems, as shown from Fig. 3, it is found that irrespective of transition metal compositions and functional groups, adsorption energies for nitride-based MXenes are lower than carbide-based MXenes. This indicates that the terminal groups adhere to the nitride-MXene surfaces more strongly than carbide-MXene surfaces.

### 3.2. Magnetic properties

Single transition metal Cr-based MXene *i.e.*, $Cr_2C$ is the first reported half-metallic ferromagnetic system amongst all the explored MXenes, which makes it an appealing candidate for nano-spintronics applications.[44] This motivated us to search for half-metallic ferromagnetic systems from the underexplored class of DTM Cr, Mo, and W based MXenes. In the periodic table, Cr, Mo, and W belong to the same group (Group-VI) and in their outer shells, they are having identical number of valence electrons. The typical oxidation number for C, N, O, F and OH are -4, -3, -2, -1 and -1 respectively and this oxidation number concept adopted on various MXenes studies previously to explain their magnetic properties.[44,45,54,71] After passivation with F, O or OH, valence d-shell of surface metal atoms (Cr/Mo/W) remain partially filled and the electrons remain localized in partially filled d-orbital. This $d^n$ configurations with $n > 0$, originate the magnetic moments in the explored systems. Calculated magnetic moments along with the stable magnetic phase (with PBE) for pristine MXenes are tabulated in Table 1 and for all the functionalized MXenes are summarized in Table S1-S3. Magnetic moments along with stable magnetic state for most stable 36 functionalized MXenes are tabulated in Table 2. From magnetic properties analysis, it has been found that majority of the explored MXenes are magnetic, and they are possessing high magnetic moments (up to 6.4 $\mu_B$ per unit cell), except Mo-W based MXenes which are nonmagnetic. For pristine STM MXenes, $Cr_3C_2$ and $Cr_3N_2$ are magnetic while Mo and W-based both STM and DTM carbide and nitride systems are nonmagnetic. To obtain the stable magnetic phase, optimized energy calculation for ferromagnetic (FM) and antiferromagnetic (AFM) states were carried out. It is found that pristine DTM carbide MXenes are mainly stable in ferromagnetic state while nitride based MXenes are stable in antiferromagnetic state, except $W_2CrC_2$ system which is stable in AFM-state. Mostly, the functionalized Cr-Mo/Cr-W based MXenes, are stable in their FM state, while $Cr_2WN_2(OH)_2$, and $W_2CrC_2F_2$ systems are stable in their AFM state. It may be noted that the value of magnetic moments for nitride based MXenes are higher than carbide based MXenes and this is because of the extra electrons introduced by nitrogen compared to carbon. The magnitude of magnetic moment also depends on the passivating groups. –F and –OH group terminated MXenes have higher magnetic moment than =O group terminated MXenes, in general. For O-group there is two unfilled orbitals while for -F and -OH group there are one unfilled orbital. As a result, more electrons assembled in surface M-centre for –F and –OH group terminated MXenes than for =O group terminated MXenes and show higher magnetic moment for –F and –OH functionalized MXenes. From the forgoing analysis it is clear that one can fine tune the magnetic behavior (NM to FM/AFM) of MXenes by selectively choosing the compositions and terminating groups.

To understand the origin of magnetic moments and contributions of different transition metal atoms on the magnetic moments, spin density distributions calculations has been performed for all magnetic systems by considering $2\times2\times1$ supercell and presented in Fig. S8-S9. From the figures, it has been observed that the net spin densities were mainly concentrated on Cr sites for both FM and AFM systems, whereas contribution of Mo/W atoms in the magnetic moment of the system is negligible. For pristine and terminated $M'_2M''X_2$ (where, $M'$= Mo/W; $M''$ = Cr; X = C/N) systems, in their AFM states, magnetic moments oriented antiparallelly to nearby Cr atoms only.

### 3.3. Electronic structure

Having studied the magnetic ground state, further, to disclose the potential applications of the most stable DTM-MXenes, predicted from structural stability analysis, the electronic structure was investigated using PBE-GGA and HSE06 functionals. PBE-GGA functional grossly underestimates the band gap due to SIE whereas HSE06 functional exhibits relatively low SIE than PBE. Although HSE06 is computationally expensive, it provides a relatively accurate band gaps and positions. Therefore, to compute the electronic structure, HSE06 was used which precisely make confidence to predict the possible applications of large set of materials under investigation. Most of the pristine MXenes tend to show metallic behavior. However, upon surface termination with chemical groups, MXenes provide a possible pathway for opening of bandgap. Based on the transition metal(s) and surface terminated groups, the corresponding materials may semi-metallic/half-metallic/semiconductor or an insulator. For instance, one can fine tune the electronic structure of MXenes from metallic to semi-metallic/half-metallic/gapless half metallic to semiconductor or even insulator by varying the surface functional groups and changing the transition metal atom(s) composition in MXene layers.[19] To ascertain the reason behind the variation of electronic structures, the spin-polarized electronic band structures were investigated methodically for pristine MXenes and their changes upon different functionalization. Spin-polarized electronic band structure analysis for all the explored systems (12 pristine and 36 stable functionalized systems) were performed using both PBE-GGA and HSE06 functionals (Fig. S1-S6). From the investigated band structures, it is found that for most cases, bands are dispersing strongly and crossing the Fermi level. This dispersing nature of band structure provides the explanation behind ferromagnetic metallic behavior of the systems. From the electronic band structure investigation of pristine DTM-MXenes, it is found that all the pristine MXenes are metallic in nature except $W_2CrC_2$ system. It is showing nearly gapless half-metallic properties with HSE06 calculation (Fig. S5(a)). The electronic band structure of –F and –OH functionalized MXenes are very similar whereas nature of band structure upon surface functionalization with =O group on pristine MXenes is different. This is caused by receiving only one electron by –F and –OH groups while =O group needs two electrons from the MXene surface for adsorption and stabilization on the surface. Most of the explored stable functionalized MXenes are showing metallic behavior while few of them possess halfmetallic or near to half-metallicity or gapless half-metallic properties.

The spin-polarized electronic band structures along with projected density of states for all Mo-W based MXenes are presented in Fig. S4 and Fig. S6. From the electronic band structures, it is clearly visualized that all the W and Mo based pristine and functionalized MXenes are metallic. $Cr_2M''X_2$ (M′′ = Mo/W; X = C/N) based pristine and all corresponding functionalized MXenes are also showing metallic behavior (Fig. S1 and Fig. S2). For $Mo_2Cr$ and $W_2Cr$ based MXenes (Fig. S3, Fig. S5), although most of the corresponding pristine and surface passivated systems are showing metallic character, two of passivated systems $Mo_2CrN_2O_2$, and $W_2CrN_2O_2$ are showing half-metallicity, while another two systems $Mo_2CrC_2O_2$, and $W_2CrC_2O_2$ found to be nearly half-metallic nature.

To verify the effect of spin–orbit coupling (SOC) on the electronic structure, we calculated SOC-included band structure for pristine $W_2CrC_2$ and functionalized $Mo_2CrC_2O_2$ and $W_2CrC_2O_2$ systems using HSE06 method and presented in Fig.S7. Generally, SOC has an important effect on 4d and 5d transition metals which have a large radius, such as Mo and W, and has little effect on 3d transition metals like Cr which have a relatively small radius. It has been found that the inclusion of SOC is not affecting much the electronic band structure and metallic nature of the considered systems.

From these large set of explored DTM complexes, specific systems which showing unique properties like half-metallicity from their electronic band structures, have been described below due to their probable potential applications in nanospintronic devices.

#### 3.3.1. Half-Metallic ferromagnetic MXenes: $Mo_2CrN_2O_2$, $W_2CrN_2O_2$

The spin-polarized electronic band structures for $Mo_2CrN_2O_2$, and $W_2CrN_2O_2$ MXenes within HSE06 are presented in Fig. 4. The most intriguing feature of the electronic band structures is the half-metallicity, i.e., the majority-spin electrons navigating a metallic behavior while a semiconducting character is observed for the minority-spin channels with a direct band gap of 0.77 eV and 1.35 eV, respectively. Thus, majority-spin channel dominated the charge transport and electric current should be completely spin-polarized for these type systems. Generally, in the low-dimensional materials like graphene, $MoS_2$, or C/BN heterostructure, halfmetallicity introduced upon doping with selective materials or by introducing strong external electric field. While the discovered $Mo_2CrN_2O_2$, and $W_2CrN_2O_2$ MXenes in the present work show intrinsic halfmetallicity.

From spin density distribution (Fig. S9) it is observed that for $Mo_2CrN_2O_2$, and $W_2CrN_2O_2$ systems, net spin concentrated on Cr-atoms and their magnetic moments oriented parallelly to nearby Cr atoms, whereas contribution of Mo/W atoms in magnetic moments are negligible.

Therefore, the ferromagnetic behaviour of $Mo_2CrN_2O_2$, and $W_2CrN_2O_2$ systems is originated from the itinerant Cr d electrons which occupy the majority-spin channel only. In $Mo_2CrN_2O_2$, and $W_2CrN_2O_2$ structure, each Cr atoms are surrounded by six N and O atoms. Thus, we assume that under $D_{3d}$ symmetry of the octahedral crystal field, degeneracy of Cr d-orbital breaks and splits the d-orbitals as low-lying $t_{2g}$ ($d_{xy}$, $d_{yz}$, $d_{xz}$) states and higher energy $e_g$ states ($d_{x^2-y^2}$, $d_{z^2}$). Due to this large exchange splitting of Cr d orbitals, 3d electrons of Cr occupy the majority spin channels and an energy gap generated between the occupied Mo/W d orbitals and vacant Cr d-orbitals along with vacant Mo/W d-orbitals in the minority spin channel (Fig.4). From the spin polarized PDOS structures of $Mo_2CrN_2O_2$, and $W_2CrN_2O_2$ system (Fig.4), it is observed that in the majority spin channel, the nature of the d orbitals are delocalized around the Fermi levels, peaks are wide, overlapping with each other and all of them are fractionally occupied. Thus, resulting the metallic behaviour for up-spin channel. Besides, from spin polarized band structure (Fig.4), it has been seen that strongly dispersing bands crossing the Fermi level for up-spin channels which also support the metallic behaviour for up-spin channel. Cr d orbitals predominantly contributing to the dispersing bands crossing the Fermi level, which supports the itinerant nature of Cr d electrons. According to Stoner theory, the itinerant d electrons favour ferromagnetism[72].

This unique physical property, the intrinsic halfmetallic ferromagnetism, makes $Mo_2CrN_2O_2$, and $W_2CrN_2O_2$ systems attractive towards applications in the field of nanoscale spintronics.

### 3.3.2. Nearly half-metallic ferromagnetic MXenes: $Mo_2CrC_2O_2$, $W_2CrC_2O_2$

Pristine $Mo_2CrX_2$ is found to be metallic for both up- and down-spin channels with ferromagnetic nature. Upon functionalization with =O group, $Mo_2CrC_2O_2$ becomes nearly half-metallic from PBE-band structure analysis whereas with HSE06 functional it is showing metallic character due to shifting of VBM towards conduction band. Similar feature has also been observed for $W_2CrC_2O_2$ system. It is also showing nearly half-metallic ferromagnetic properties with PBE-band structure analysis and metallic behavior with HSE06 method. The obtained spin-polarized electronic band structures and projected density of states with PBE and HSE06 for $Mo_2CrC_2O_2$ and $W_2CrC_2O_2$ systems are provided in Fig. 5.

### 3.4. Lattice Dynamics

To ensure the dynamic stability of the predicted ground-state structures for the MXenes calculated within PBE which possess striking features like, ferromagnetic intrinsic half-metallic/near to half-metallic properties, lattice dynamical calculations were carried out to compute phonon dispersion curves. For $Mo_2CrC_2O_2$ ($H_XH_X$) system, the lattice dynamical calculations were performed with 2x2x1, 3x3x1, 4x4x1, and 5x5x1 supercells with k-mesh of 6x6x1, 3x3x1, 2x2x1, and 2x2x1, respectively, as shown in Fig S10. From the figure it has been observed that 2x2x1 supercell resulting positive phonon frequencies, whereas for 3x3x1 supercell we could see nearly stable or noise in the acoustic transverse phonons, which improved for 4x4x1 supercell. Finally, 5x5x1 supercell provide positive phonon frequencies along with quadratic behaviours of transverse acoustic phonons which is a typical behaviour of a 2D material. Hence, 5x5x1 supercell was used to compute phonon dispersion curves along the high symmetry direction of the Brillouin zone (BZ) for $Mo_2CrC_2O_2$ ($H_XH_X$), $Mo_2CrN_2O_2$ ($H_XH_X$), $W_2CrC_2O_2$ ($H_XH_X$), and $W_2CrN_2O_2$ ($H_XH_X$) stable MXenes and are presented in Fig. 6. Since, the unit cell consists of 7 atoms, it is resulting 21 phonon modes, out of which 3 are acoustic and 18 are optical phonon modes. As shown in Fig. 6, all the positive phonon frequencies throughout the BZ indicating the dynamic stability of the predicted ground-state structure of these MXenes for possible synthesis in near future.

## 4. Conclusions

The euphoria associated with understanding the physics and chemistry of inorganic 2D materials has been at its peak in recent years, owing to their unique properties and multifunctional behavior. In this work, a comprehensive study has been carried out on the surface terminated double transition metal-based carbide and nitride MXenes ($M'_2M''X_2\ T_x$; M′ and M′′ = Cr, Mo, W; X = C or N; T = –F, –OH, or =O) (144 configurations) using first-principles calculations based on spin-polarized DFT. Stable and potential candidate materials were screened from large set of explored systems for possible device applications. The calculated total and adsorption energies were employed to assess the thermodynamic stability of 144 functionalized MXenes with four possible models and screened 36 stable MXene structures, provides insight on the feasibility of synthesis. Spin-resolved electronic structure calculations were performed on these 36 stable systems using both PBE-GGA and HSE06 functionals. The stable MXenes exhibit wide range of magnetic ground states from non-magnetic to ferromagnetic, and then to antiferromagnetic. All the Mo and W-based both carbide and nitride MXene systems are non-magnetic and metallic whereas Cr-Mo and Cr-W based carbide and nitride pristine and surface passivated MXenes are metals/half metals/gapless half-metallic with relatively high magnetic moments (up to 6.4 $\mu_B$ per unit cell) and stable in ferromagnetic state for most of the ground state structure except $Cr_2WN_2(OH)_2$, $W_2CrC_2F_2$ systems. They are stable in their AFM state. From the spin-polarized electronic band structure analysis, it is predicted that $Mo_2CrN_2O_2$, and $W_2CrN_2O_2$ systems are half-metallic ferromagnetic materials. They are showing metallic nature for major-spin channel and semiconducting for minor-spin channels with a direct band gap of 0.77eV, and 1.35eV, respectively. For $Mo_2CrC_2O_2$, and $W_2CrC_2O_2$ system in PBE method it is found that they are nearly halfmetallic ferromagnetic materials with a band gap 0.13eV and 0.18eV respectively for down spin electrons while showing metallic behavior upon analysis with HSE06 functional. Besides, pristine $W_2CrC_2$ is a gapless antiferromagnetic material.

Among the explored MXenes so far in the literature, the number of compounds which have intrinsic magnetism and halfmetallic property are scarce. Therefore, computational screening of a fairly large set of possible Cr, Mo, and W-based DTM carbide and nitride MXenes, narrows down to a few suitable candidate materials which have unique intrinsic ferromagnetism/antiferromagnets half metallic/gapless half-metallic properties, makes them promising materials for possible synthesis and their potential applications in the field of nanoscale spintronic and electronic devices.

## Conflicts of interest

There are no conflicts to declare.

## Author Contributions

The manuscript is written through the contribution of all authors. All authors have given approval to the final version of the manuscript.

## Acknowledgments

The authors gratefully acknowledge funding by DST-SERB-EEQ [EEQ/2019/000736] and DST-Nanomission [DST/NM/NS/2018/204(G)], New Delhi, India for financial support. G. N. Sastry thanks DST for JC Bose fellowship. The authors thank CSIR-NEIST, Jorhat, India, for providing the necessary computational facilities and support.

## Supporting Information

Electronic spin-polarized and non-spin-polarized band structure; Projected density of states (PDOS); Spin orbit coupling included electronic band structure (HSE06+SOC); Magnetic spin density distributions; Phonon dispersion calculations; Relative energies (with and without dispersion correction), and magnetic moment comparison with four modelled structures (Model I-IV) based on probable passivation sites on DTM MXenes surface considering different functional groups; Typical input files for geometry optimization and phonon dispersion calculations, Optimized geometries for 36 stable MXenes and 4 MXenes used for lattice dynamical calculations with VASP.

## Tables and Figures

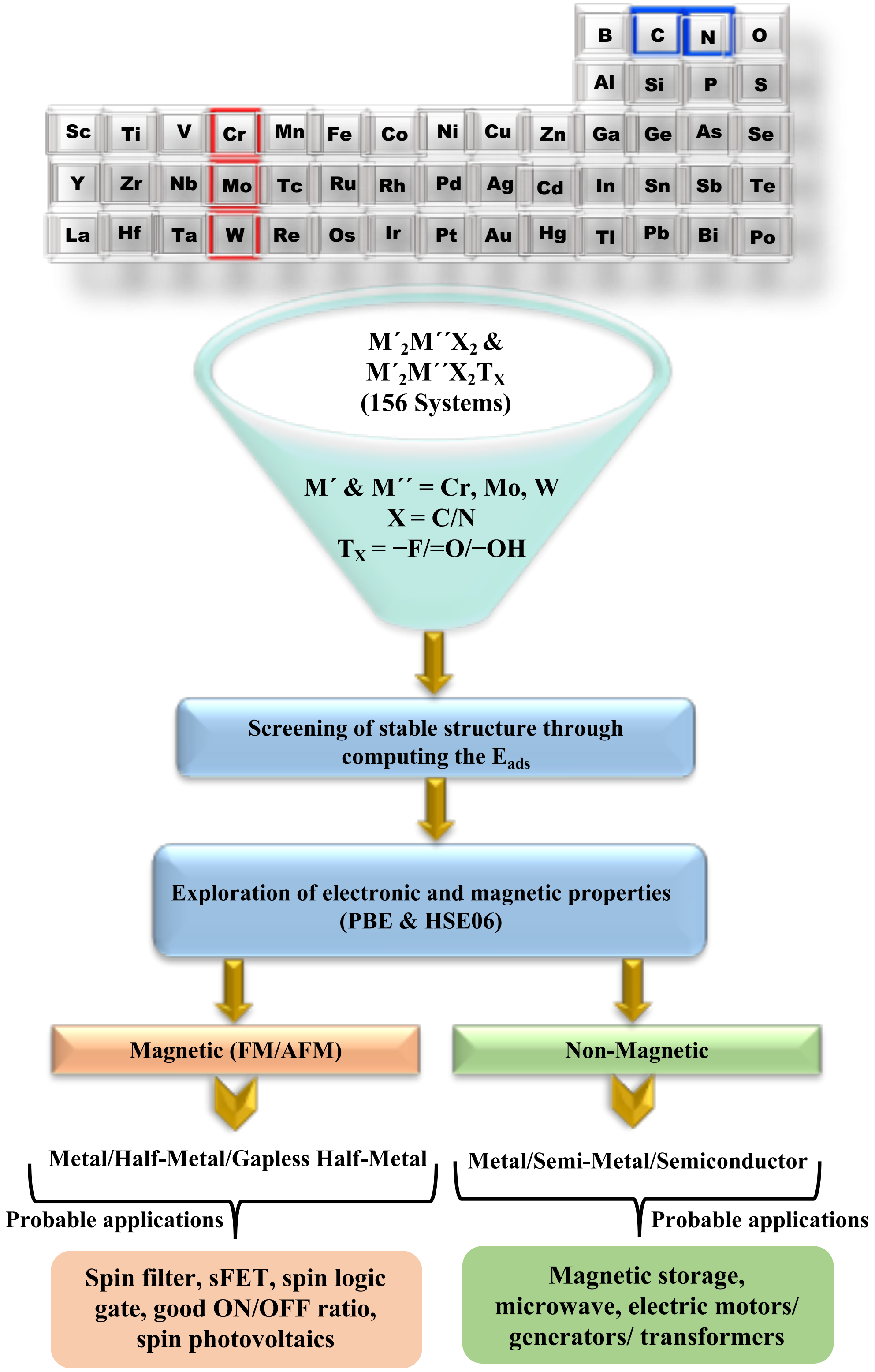

**Fig. 1:** Workflow for screening stable functionalized MXenes, exploration of magnetic properties and electronic structures and their possible potential applications.

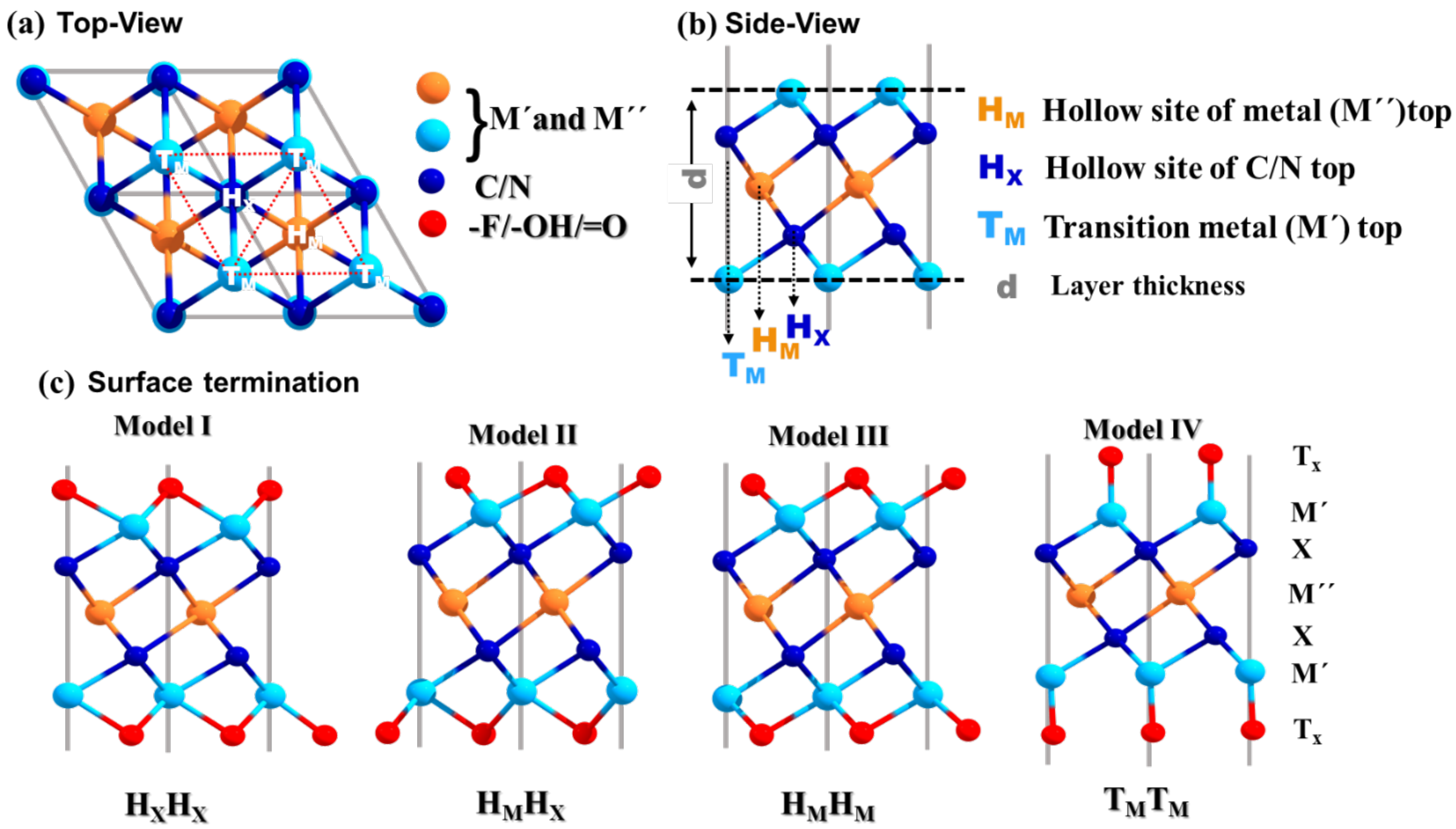

**Fig. 2:** Representative structures of pristine $M'_2M''X_2$ MXenes (a) top view, (b) side view. Side views of (c) Model I-IV for functionalized $M'_2M''X_2T_x$ MXenes (M´ and M´´= Cr/Mo/W; X = C/N; T = –F/–OH/=O).

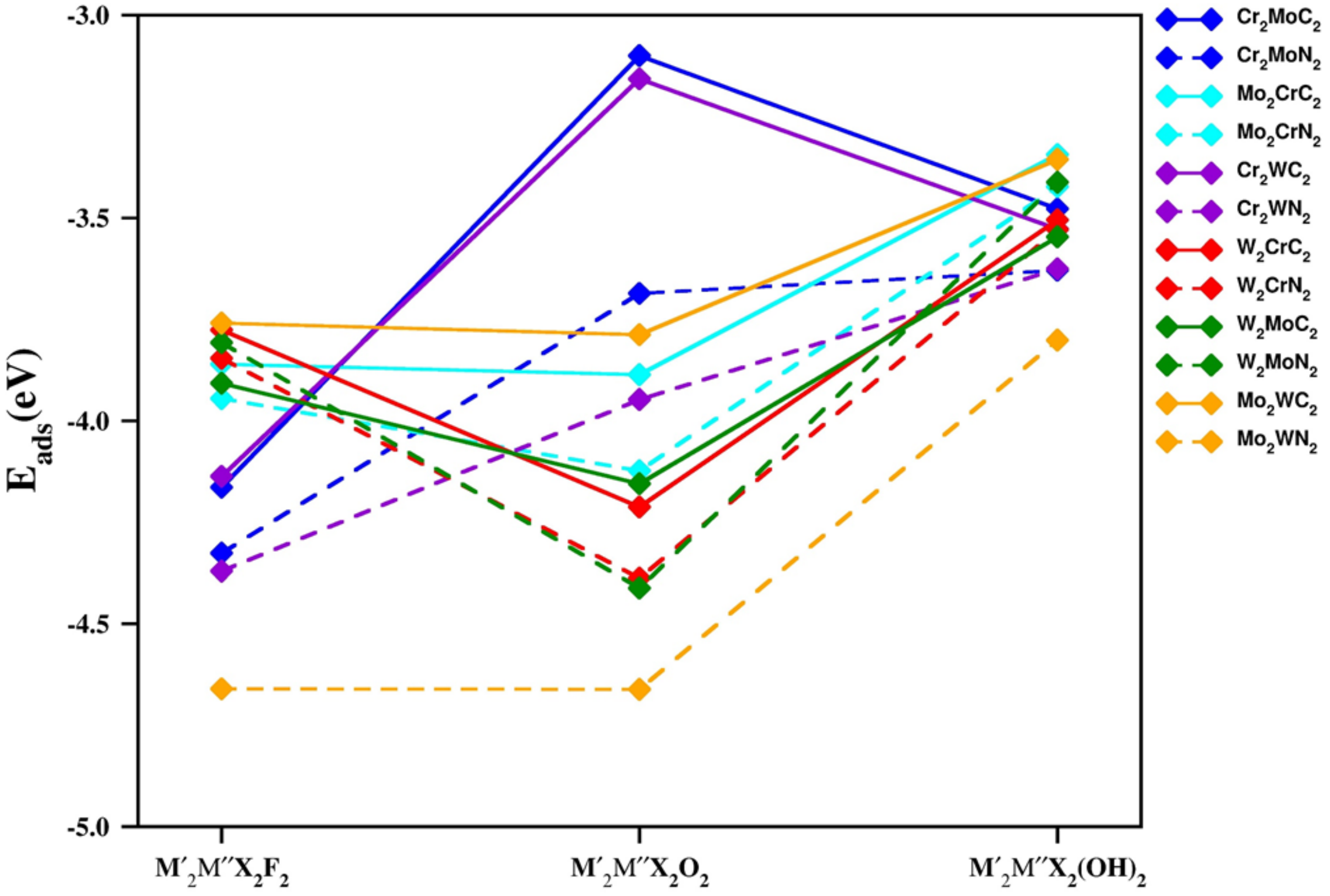

**Fig. 3:** The adsorption energies ($E_{ads}$) for adhesion of terminal groups (–F, =O, and –OH) onto the surfaces of M′$_2$M′′$X_2$ systems (M′ and M′′ = Cr, Mo, W; X = C/N).

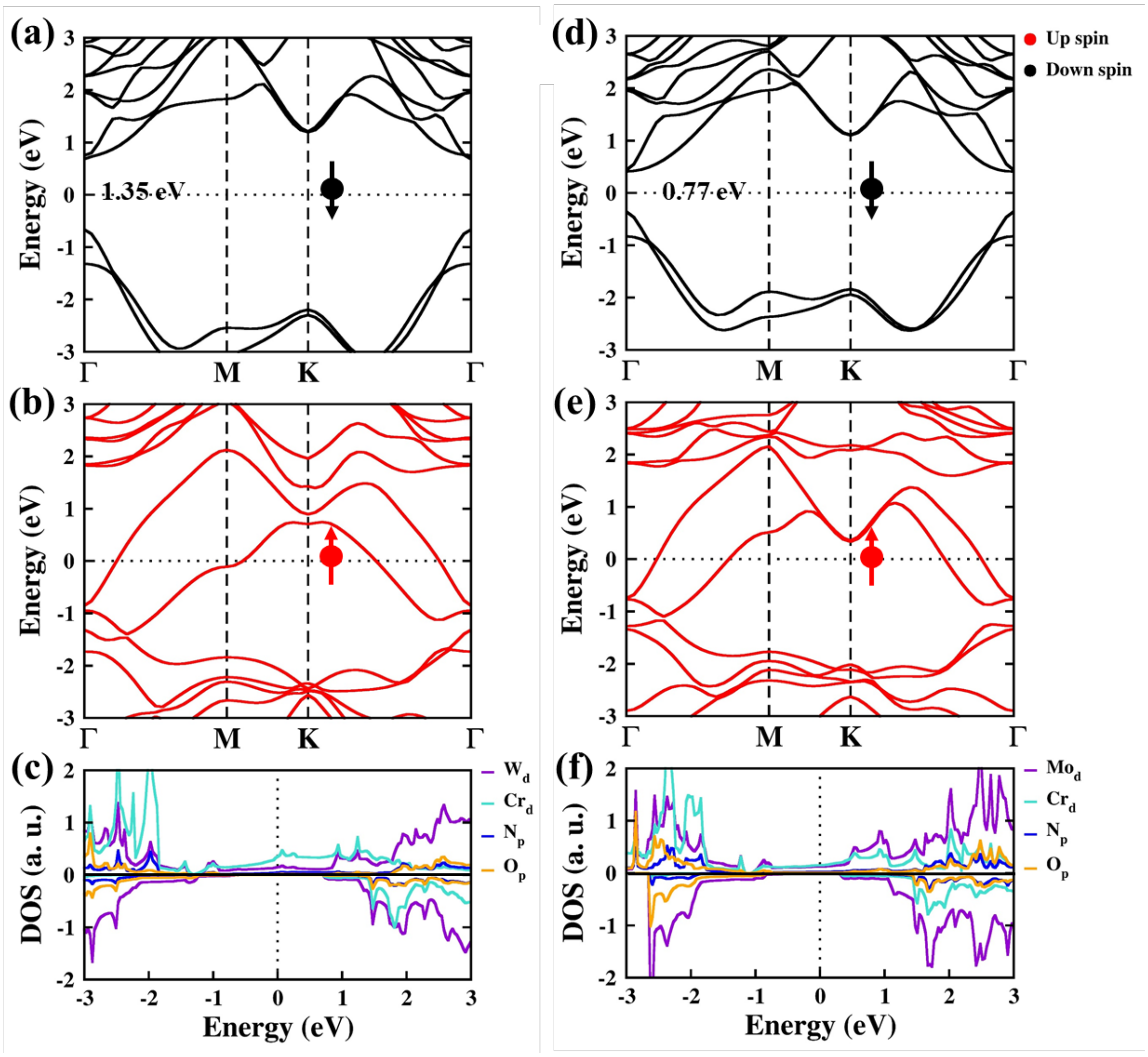

**Fig. 4.** Electronic spin-polarized band structure for (a) down-spin, (b) up-spin, and (c) spin-resolved projected density of states (PDOS) with HSE06 functional for $W_2CrN_2O_2$ system**.** Electronic spin-polarized band structure for (d) down-spin, (e) up-spin and (f) spin-resolved projected density of states (PDOS) with HSE06 functional for $Mo_2CrN_2O_2$ system. The Fermi energy is shifted to zero and indicated by the horizontal dashed black line. For down-spin semiconducting systems, fermi energy is shifted to the centre of the gap.

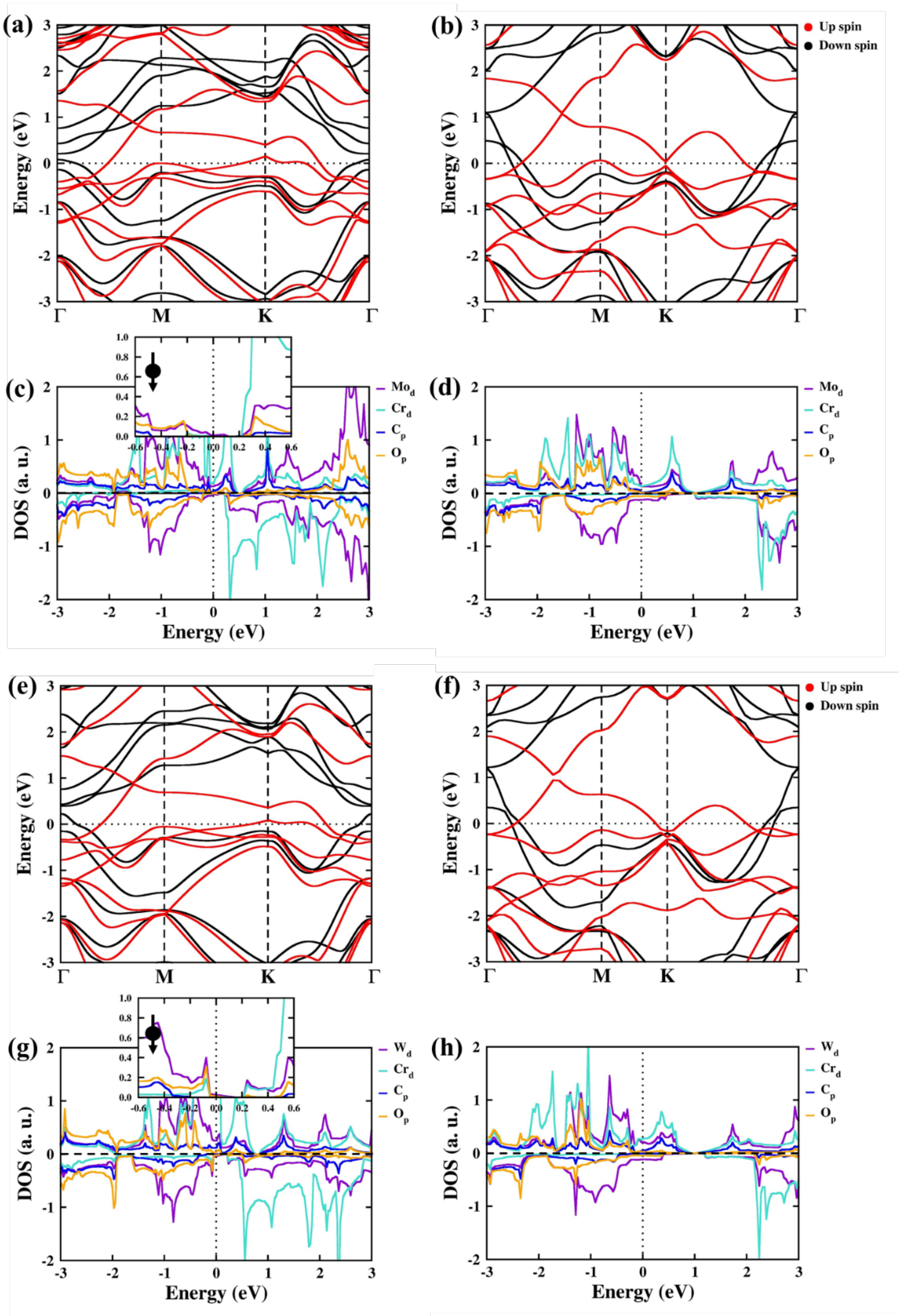

**Fig. 5.** Electronic spin-polarized band structure (a) with PBE functional, (b) with HSE06 functional, and spin-resolved projected density of states (c) with PBE functional, (d) with HSE06 functional for $Mo_2CrC_2O_2$ system. For $W_2CrC_2O_2$ system, electronic spin-polarized

band structure (e) with PBE functional, (f) with HSE06 functional, and spin-resolved projected density of states (g) with PBE functional, (h) with HSE06 functional. The Fermi energy is shifted to zero and indicated by the horizontal dashed black line. The insets in (c) and (g) have enlarged PDOS plot around the Fermi level for down-spin channel.

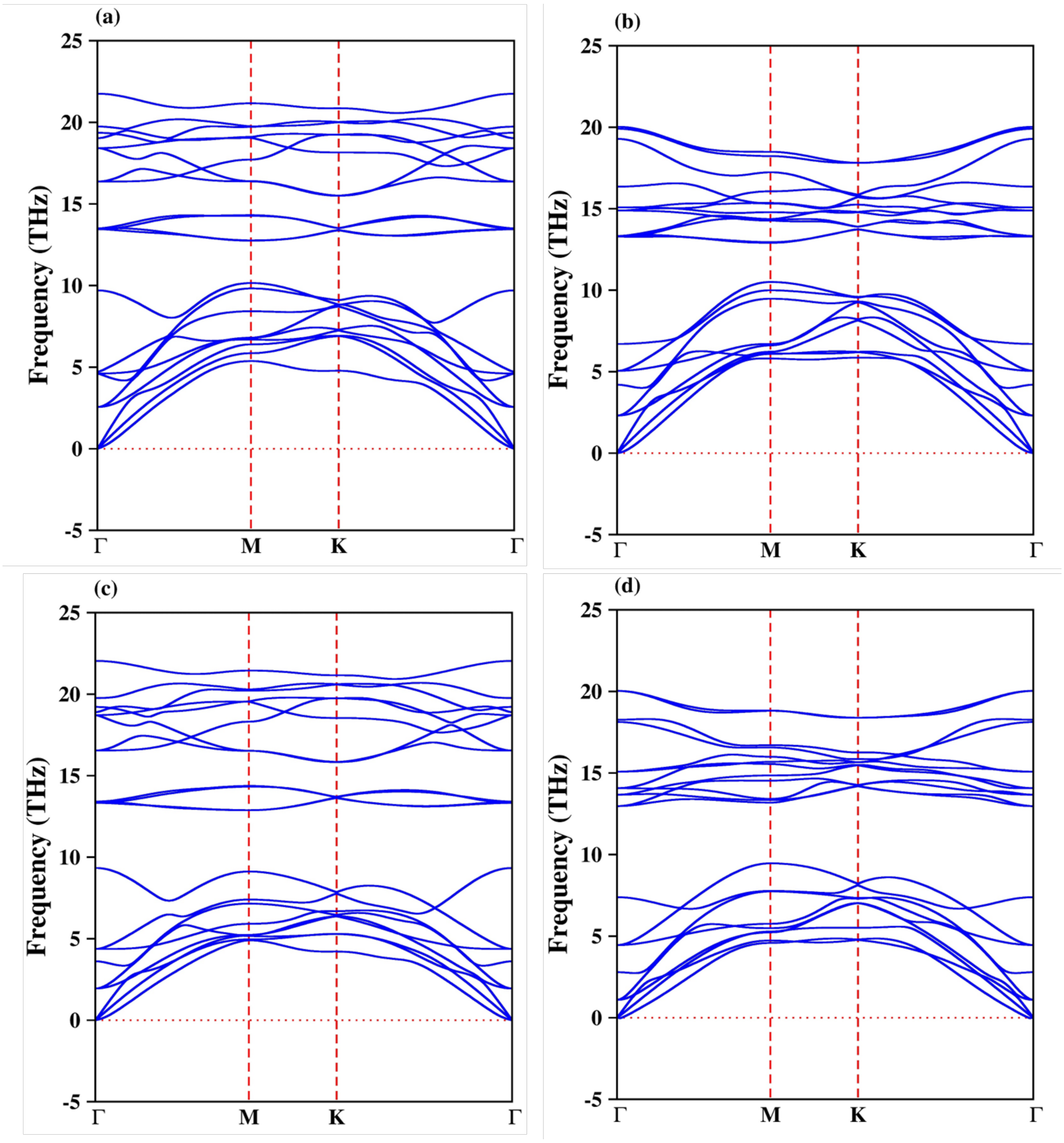

**Fig. 6:** The phonon dispersion calculated within PBE for (a) $Mo_2CrC_2O_2$ ($H_XH_X$), (b) $Mo_2CrN_2O_2$ ($H_XH_X$), (c) $W_2CrC_2O_2$ ($H_XH_X$), and (d) $W_2CrN_2O_2$ ($H_XH_X$) MXenes with 5x5x1 supercell. $H_XH_X$ indicating, hollow site of C/N top for top and bottom surface of MXenes.

**Table 1:** Calculated lattice constant *a* (Å), inter-layer distance *d* (Å) without (w/o) and with vdW dispersion correction, normalized cohesive energy $\bar{E}_{coh}$ (eV/atom), magnetic moment at ferromagnetic state (MM in $\mu_B$/unit cell), and magnetic energy difference ΔE = $E_{AFM}$-$E_{FM}$ (eV) of double transition metal based pristine MXenes ($M'_2M''X_2$; M′ and M′′ = Cr, Mo, W; X = C/N), in comparison with single transition metal based MXenes ($M_3X_2$; M = Cr/Mo/W; X = C/N).

| **Pristine MXene** | ***a*** | ***d*** | | $\bar{E}_{coh}$ | **MM** | **ΔE** |
|---|---|---|---|---|---|---|
| | | **M′-X w/o vdW (M′-X with vdW)** | **M′′-X w/o vdW (M′′-X with vdW)** | | | |
| $Cr_3C_2$ | 3.02 | 1.93 (1.92) | 2.12 (2.11) | -5.15 | 5.42 | 0.112 |
| $Cr_3N_2$ | 2.98 | 1.89 (1.92) | 2.15 (2.11) | -4.35 | 4.47 | 0.530 |
| $Cr_2MoC_2$ | 3.09 | 1.95 (1.94) | 2.25 (2.24) | -5.68 | 5.54 | 0.303 |
| $Cr_2MoN_2$ | 3.06 | 1.89 (1.88) | 2.27 (2.26) | -4.81 | 4.27 | -0.054 |
| $Cr_2WC_2$ | 3.09 | 1.96 (1.95) | 2.24 (2.23) | -6.06 | 5.49 | 0.425 |
| $Cr_2WN_2$ | 3.07 | 1.91 (1.90) | 2.22 (2.23) | -5.08 | 4.31 | -0.438 |
| $Mo_3C_2$ | 3.02 | 2.06 (2.06) | 2.19 (2.19) | -6.47 | 0.00 | — |
| $Mo_3N_2$ | 2.82 | 2.13 (2.12) | 2.20 (2.19) | -5.60 | 0.00 | — |
| $Mo_2CrC_2$ | 2.94 | 2.07 (2.07) | 2.08 (2.06) | -6.00 | 1.15 | 0.041 |
| $Mo_2CrN_2$ | 2.87 | 2.11 (2.10) | 2.08 (2.06) | -5.14 | 2.73 | -0.652 |
| $Mo_2WC_2$ | 2.98 | 2.07 (2.07) | 2.19 (2.18) | -6.85 | 0.00 | — |
| $Mo_2WN_2$ | 3.09 | 2.07 (2.07) | 2.16 (2.15) | -5.61 | 0.00 | — |
| $W_3C_2$ | 3.00 | 2.08 (2.08) | 2.18 (2.17) | -7.55 | 0.00 | — |
| $W_3N_2$ | 2.79 | 2.15 (2.14) | 2.22 (2.22) | -6.69 | 0.00 | — |
| $W_2CrC_2$ | 2.91 | 2.07 (2.08) | 2.07 (2.05) | -6.74 | 0.01 | -0.007 |
| $W_2CrN_2$ | 2.84 | 2.12 (2.12) | 2.09 (2.07) | -5.89 | 3.01 | -0.796 |
| $W_2MoC_2$ | 2.98 | 2.07 (2.07) | 2.19 (2.18) | -7.19 | 0.00 | — |
| $W_2MoN_2$ | 2.80 | 2.09 (2.11) | 2.07 (2.08) | -6.26 | 0.00 | — |

**Table 2:** Calculated stable adsorption site, adsorption energies ($E_{ads}$), magnetic moment at ferromagnetic state (MM in $\mu_B$/unit cell), magnetic energy difference $\Delta E = E_{AFM}-E_{FM}$ (eV), and metallic behavior of the functionalized double transition metal based carbide and nitride MXenes ($M'_2M''X_2T_x$; M′ and M′′ = Cr, Mo, W; X = C/N; T = –F/–OH/=O).

| System | Stable adsorption site | $E_{ads}$ | MM | ΔE | Metallic behavior (GGA-PBE) | Metallic behavior (HSE06) |
|---|---|---|---|---|---|---|
| **$Cr_2MoC_2F_2$** | $H_XH_X$ | -4.16 | 4.69 | 1.195 | Metallic | Metallic |
| **$Cr_2MoC_2O_2$** | $H_MH_M$ | -3.10 | 4.24 | 0.041 | Metallic | Metallic |
| **$Cr_2MoC_2(OH)_2$** | $H_MH_M$ | -3.48 | 4.58 | 1.018 | Metallic | Metallic |
| **$Cr_2MoN_2F_2$** | $H_MH_X$ | -4.33 | 6.31 | 1.073 | Metallic | Metallic |
| **$Cr_2MoN_2O_2$** | $H_MH_M$ | -3.69 | 5.39 | 1.044 | Metallic | Metallic |
| **$Cr_2MoN_2(OH)_2$** | $H_MH_M$ | -3.63 | 6.17 | 0.867 | Metallic | Metallic |
| **$Cr_2WC_2F_2$** | $H_MH_X$ | -4.14 | 4.71 | 1.209 | Metallic | Metallic |
| **$Cr_2WC_2O_2$** | $H_MH_M$ | -3.16 | 4.27 | 0.834 | Metallic | Metallic |
| **$Cr_2WC_2(OH)_2$** | $H_MH_M$ | -3.53 | 4.65 | 0.981 | Metallic | Metallic |
| **$Cr_2WN_2F_2$** | $H_MH_X$ | -4.37 | 5.60 | 0.867 | Metallic | Metallic |
| **$Cr_2WN_2O_2$** | $H_MH_M$ | -3.95 | 5.65 | 1.177 | Metallic | Metallic |
| **$Cr_2WN_2(OH)_2$** | $H_MH_M$ | -3.90 | 5.91 | -0.099 | Metallic | Metallic |
| **$Mo_2CrC_2F_2$** | $H_MH_X$ | -3.86 | 1.78 | 0.178 | Metallic | Metallic |
| **$Mo_2CrC_2O_2$** | $H_XH_X$ | -3.89 | 2.02 | 0.178 | Nearly Halfmetallic | Metallic |
| **$Mo_2CrC_2(OH)_2$** | $H_XH_X$ | -3.34 | 0.00 | – | Metallic | Metallic |
| **$Mo_2CrN_2F_2$** | $H_MH_X$ | -3.95 | 2.92 | 0.502 | Metallic | Metallic |
| **$Mo_2CrN_2O_2$** | $H_XH_X$ | -4.12 | 3.54 | 0.144 | Nearly Halfmetallic | Halfmetallic |
| **$Mo_2CrN_2(OH)_2$** | $H_MH_M$ | -3.42 | 2.83 | 0.599 | Metallic | Metallic |
| **$Mo_2WC_2F_2$** | $H_XH_X$ | -3.76 | 0.00 | – | Metallic | Metallic |
| **$Mo_2WC_2O_2$** | $H_XH_X$ | -3.79 | 0.00 | – | Metallic | Metallic |
| **$Mo_2WC_2(OH)_2$** | $T_MT_M$ | -3.36 | 0.00 | – | Metallic | Metallic |
| **$Mo_2WN_2F_2$** | $H_XH_X$ | -4.66 | 0.00 | – | Metallic | Metallic |
| **$Mo_2WN_2O_2$** | $H_XH_X$ | -4.66 | 0.00 | – | Metallic | Metallic |
| **$Mo_2WN_2(OH)_2$** | $T_MT_M$ | -3.80 | 0.00 | – | Metallic | Metallic |
| **$W_2CrC_2F_2$** | $T_MT_M$ | -3.78 | 1.19 | -0.051 | Metallic | Metallic |
| **$W_2CrC_2O_2$** | $H_XH_X$ | -4.21 | 2.03 | 0.158 | Nearly Halfmetallic | Metallic |
| **$W_2CrC_2(OH)_2$** | $T_MT_M$ | -3.50 | 1.32 | 0.096 | Metallic | Metallic |
| **$W_2CrN_2F_2$** | $T_MT_M$ | -3.85 | 2.75 | 0.492 | Metallic | Metallic |
| **$W_2CrN_2O_2$** | $H_XH_X$ | -4.39 | 3.56 | 0.153 | Nearly Halfmetallic | Halfmetallic |
| **$W_2CrN_2(OH)_2$** | $T_MT_M$ | -3.53 | 2.72 | 0.598 | Metallic | Metallic |
| **$W_2MoC_2F_2$** | $T_MT_M$ | -3.91 | 0.00 | – | Metallic | Metallic |
| **$W_2MoC_2O_2$** | $H_XH_X$ | -4.15 | 0.00 | – | Metallic | Metallic |
| **$W_2MoC_2(OH)_2$** | $T_MT_M$ | -3.55 | 0.00 | – | Metallic | Metallic |
| **$W_2MoN_2F_2$** | $T_MT_M$ | -3.84 | 0.00 | – | Metallic | Metallic |
| **$W_2MoN_2O_2$** | $H_XH_X$ | -4.44 | 0.00 | – | Metallic | Metallic |
| **$W_2MoN_2(OH)_2$** | $T_MT_M$ | -3.44 | 0.00 | – | Metallic | Metallic |

## Supporting Information

# 2D-Double Transition Metal MXenes for Spintronics Applications: Surface Functionalization Induced Ferromagnetic Half-Metallic Complexes

Kripa Dristi Dihingia[a,b], Swagata Saikia[a], N. Yedukondalu[a], Supriya Saha[a,b,*], G. Narahari Sastry[a,b]

[a]Advanced Computation and Data Sciences Division, Council of Scientific and Industrial Research-North East Institute of Science and Technology (CSIR-NEIST), Jorhat, 785006, Assam, India.

[b]Academy of Scientific and Innovative Research (AcSIR), Ghaziabad, 201002, Uttar Pradesh, India.

*E-mail: supriya.saha@neist.res.in

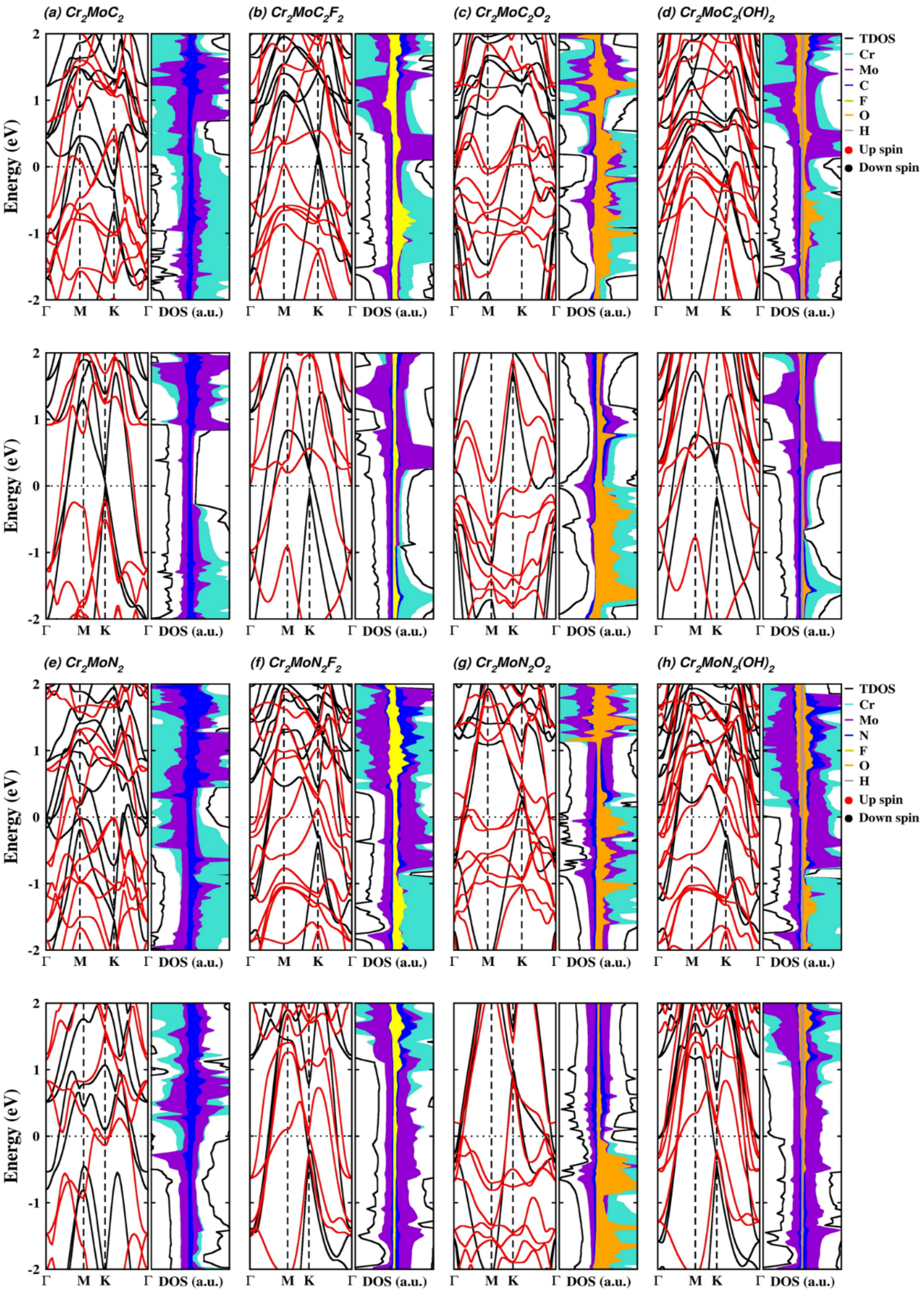

**Fig. S1:** Electronic spin-polarized band structures along with projected density of states (PDOS) of pristine $Cr_2MoX_2$ and functionalized $Cr_2MoX_2T_2$ (where X = C/N, and T = –F/–OH/=O) MXenes. Top row (PBE) and 2nd row (HSE06) for pristine and functionalized $Cr_2MoC_2$ systems [(a)-(d)]. 3rd row (PBE), and bottom row (HSE06) for pristine and functionalized $Cr_2MoN_2$ systems [(e)-(h)]. The Fermi energy is shifted to zero and indicated by the horizontal dashed black line.

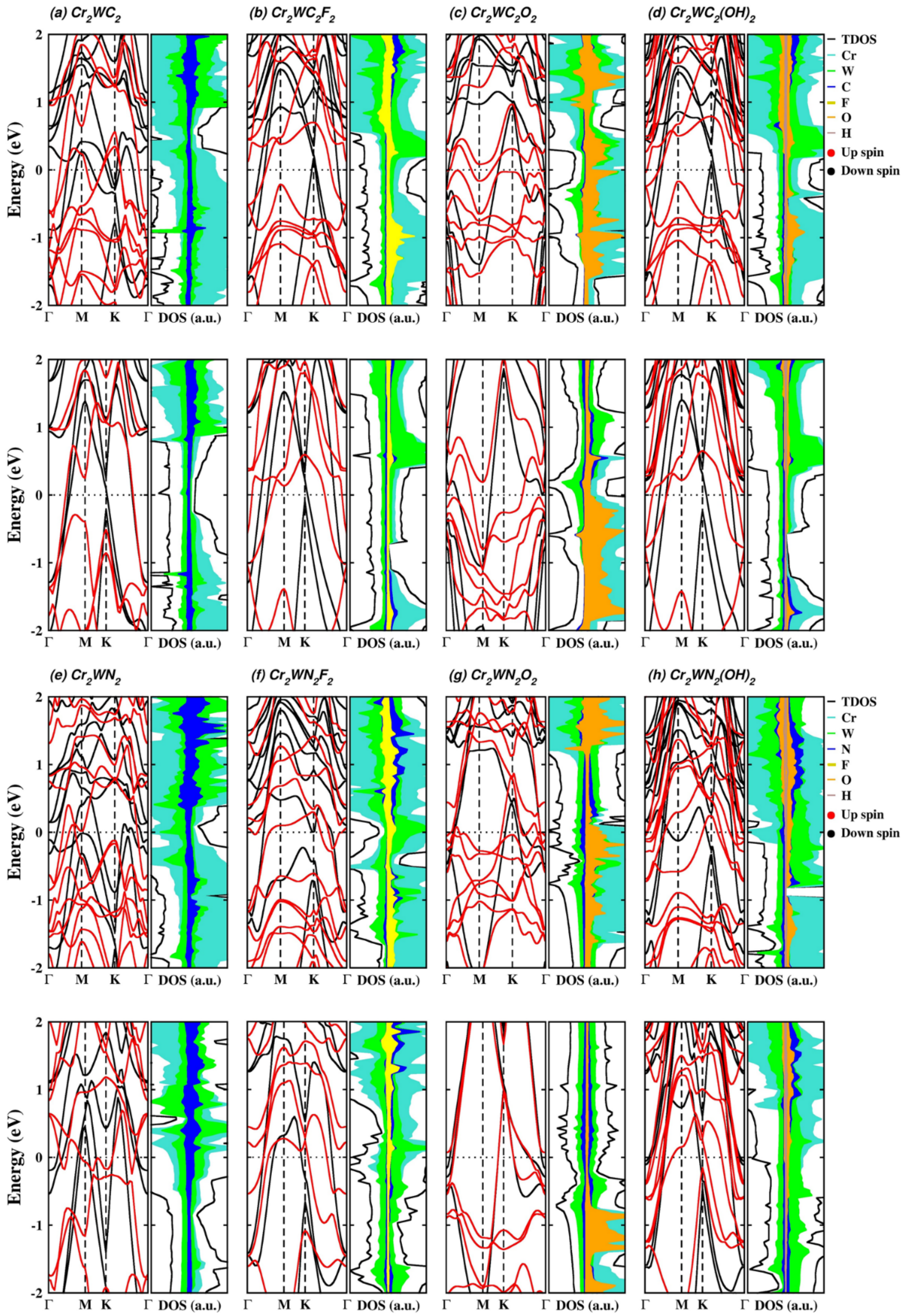

**Fig. S2:** Electronic spin-polarized band structures and projected density of states (PDOS) of pristine $Cr_2WX_2$ and functionalized $Cr_2WX_2T_2$ (where X = C/N, and T = –F/–OH/=O) MXenes. Top row (PBE) and 2nd row (HSE06) for pristine and functionalized $Cr_2WC_2$ systems [(a)-(d)]. 3rd row (PBE), and bottom row (HSE06) for pristine and functionalized $Cr_2WN_2$ systems [(e)-(h)]. The Fermi energy is shifted to zero and indicated by the horizontal dashed black line.

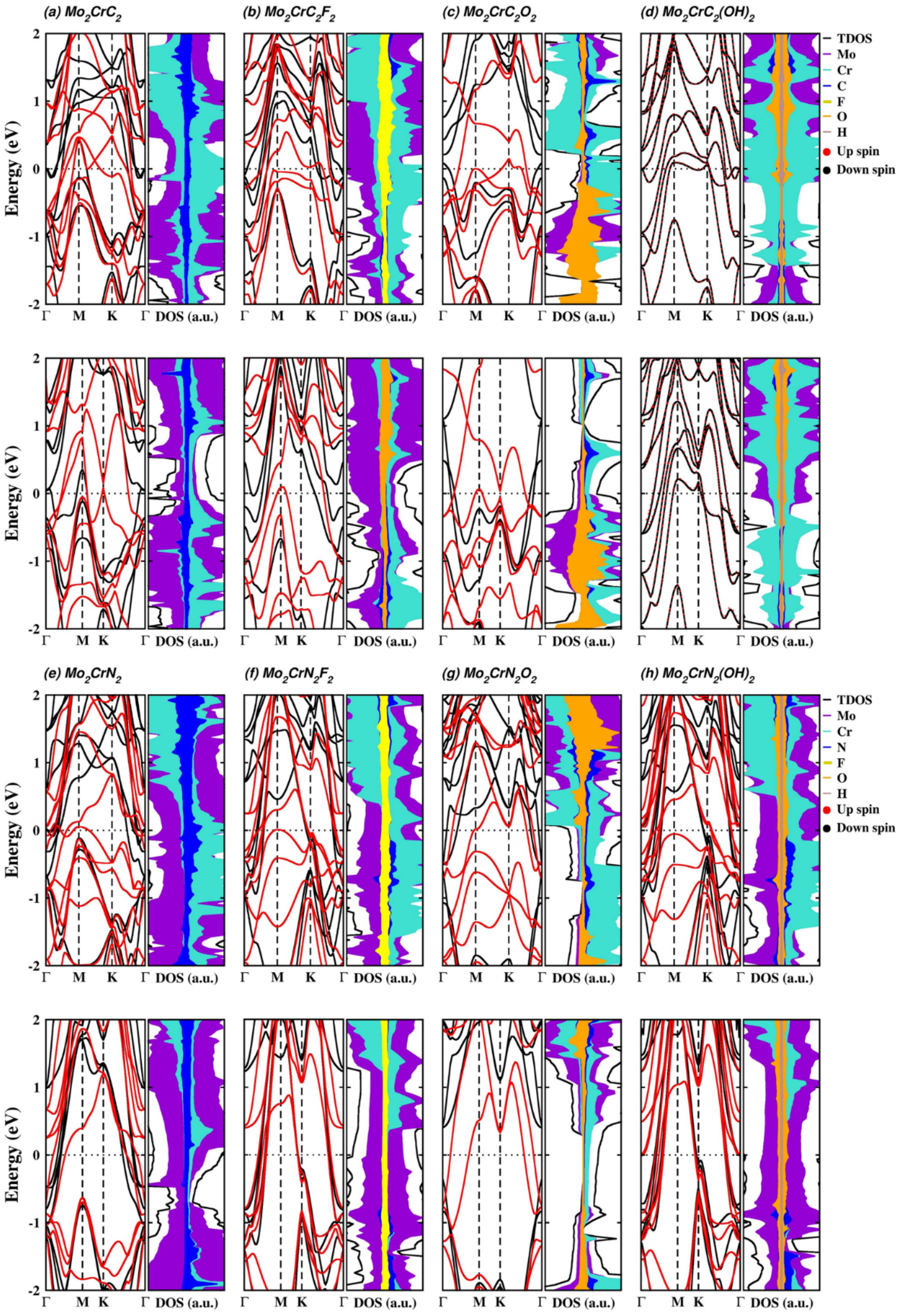

**Fig. S3:** Electronic spin-polarized band structures and projected density of states (PDOS) of pristine $Mo_2CrX_2$ and functionalized $Mo_2CrX_2T_2$ (where X = C/N, and T = –F/–OH/=O) MXenes. Top row (PBE) and 2[nd] row (HSE06) for pristine and functionalized $Mo_2CrC_2$ systems [(a)-(d)]. 3[rd] row (PBE), and bottom row (HSE06) for pristine and functionalized $Mo_2CrN_2$ systems [(e)-(h)]. The Fermi energy is shifted to zero and indicated by the horizontal dashed black line.

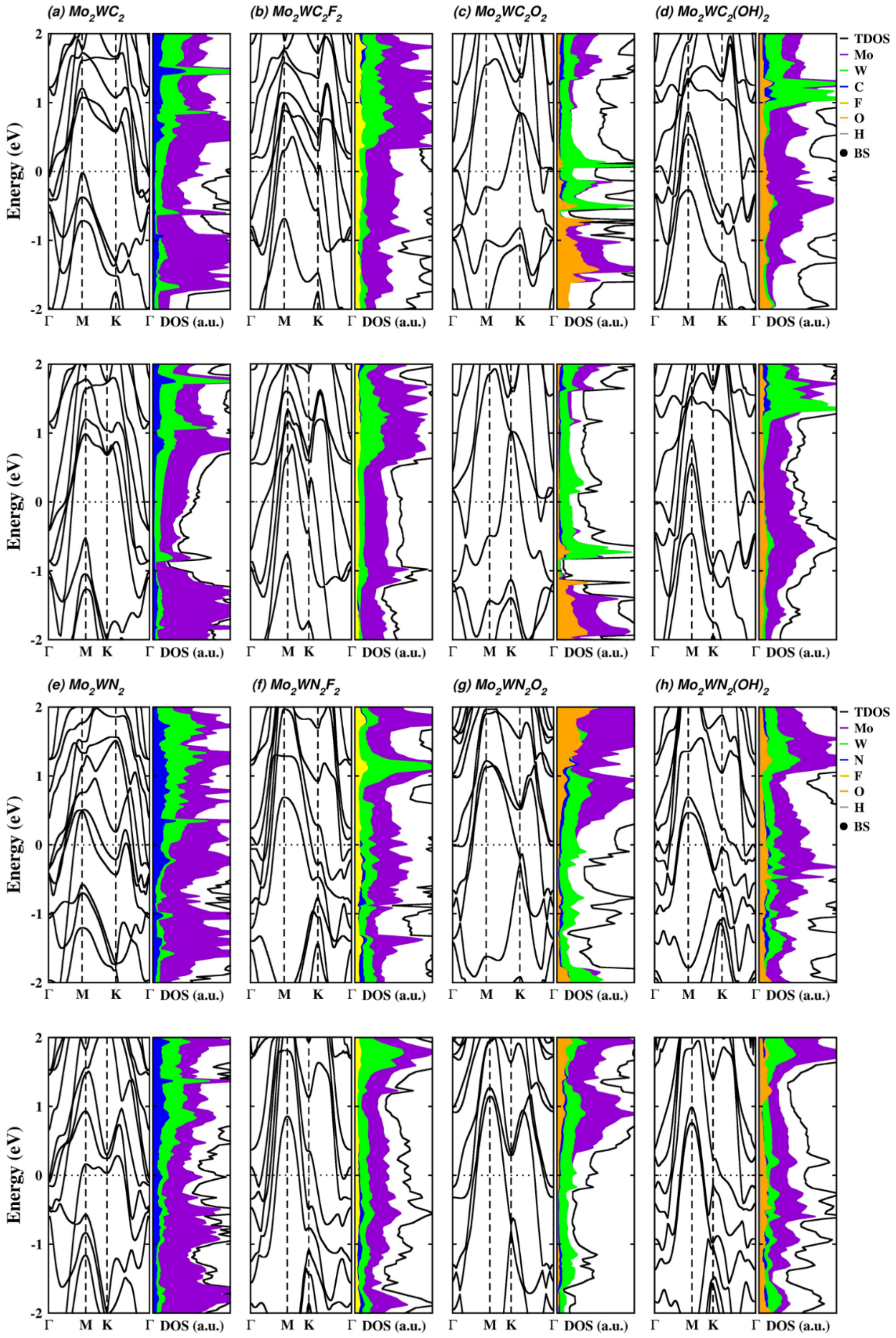

**Fig. S4:** Electronic spin-polarized band structures and projected density of states (PDOS) of pristine $Mo_2WX_2$ and functionalized $Mo_2WX_2T_2$ (where X = C/N, and T = –F/–OH/=O) MXenes. Top row (PBE) and 2nd row (HSE06) for pristine and functionalized $Mo_2WC_2$ systems [(a)-(d)]. 3rd row (PBE), and bottom row (HSE06) for pristine and functionalized $Mo_2WN_2$ systems [(e)-(h)]. The Fermi energy is shifted to zero and indicated by the horizontal dashed black line.

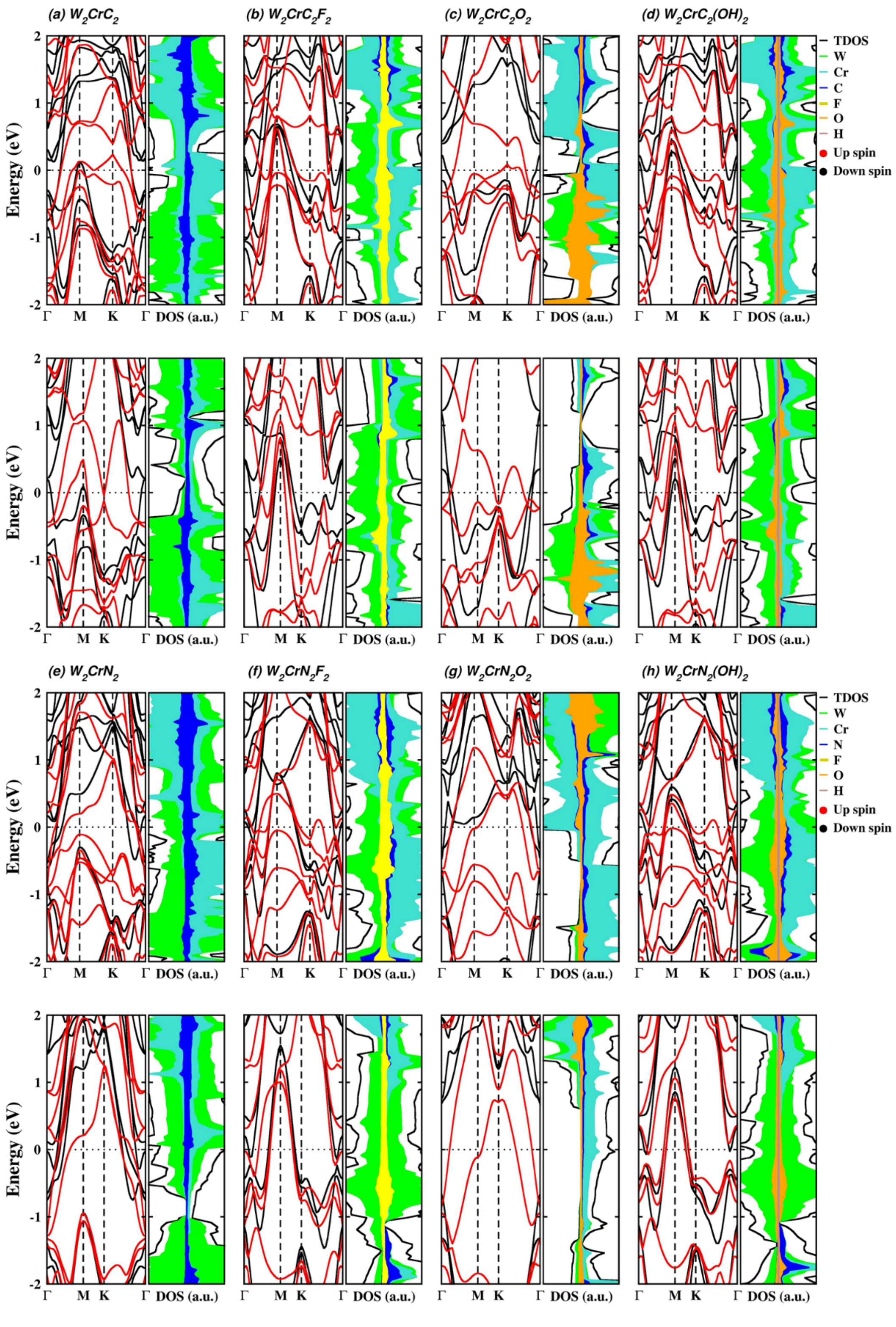

**Fig. S5:** Electronic spin-polarized band structures and projected density of states (PDOS) of pristine $W_2CrX_2$ and functionalized $W_2CrX_2T_2$ (where X = C/N, and T = –F/–OH/=O) MXenes. Top row (PBE) and 2nd row (HSE06) for pristine and functionalized $W_2CrC_2$ systems [(a)-(d)]. 3rd row (PBE), and bottom row (HSE06) for pristine and functionalized $W_2CrN_2$ systems [(e)-(h)]. The Fermi energy is shifted to zero and indicated by the horizontal dashed black line.

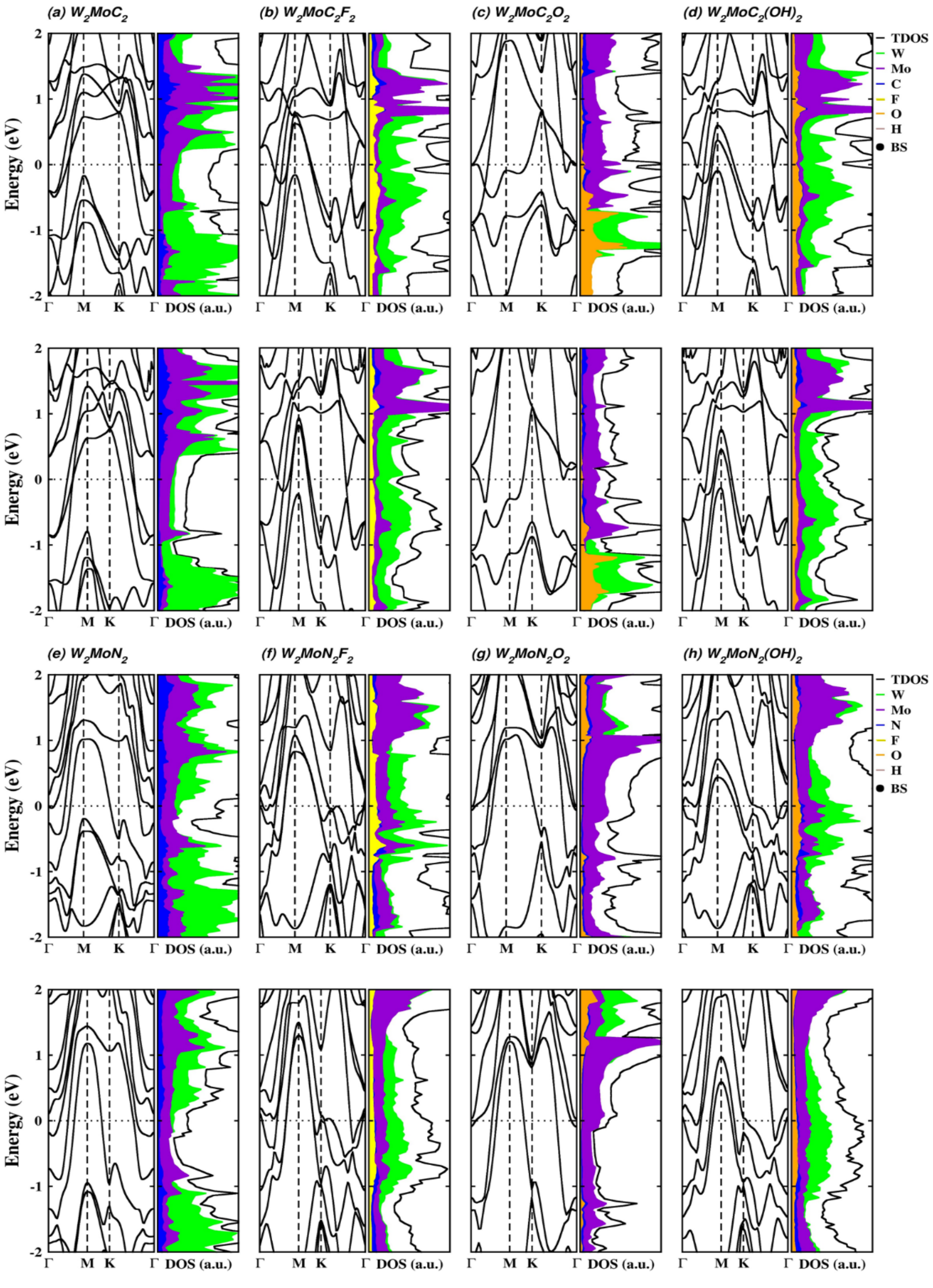

**Fig. S6:** Electronic spin-polarized band structures and projected density of states (PDOS) of pristine $W_2MoX_2$ and functionalized $W_2MoX_2T_2$ (where X = C/N, and T = –F/–OH/=O) MXenes. Top row (PBE) and 2nd row (HSE06) for pristine and functionalized $W_2MoC_2$ systems [(a)-(d)]. 3rd row (PBE), and bottom row (HSE06) for pristine and functionalized $W_2MoN_2$ systems [(e)-(h)]. The Fermi energy is shifted to zero and indicated by the horizontal dashed black line.

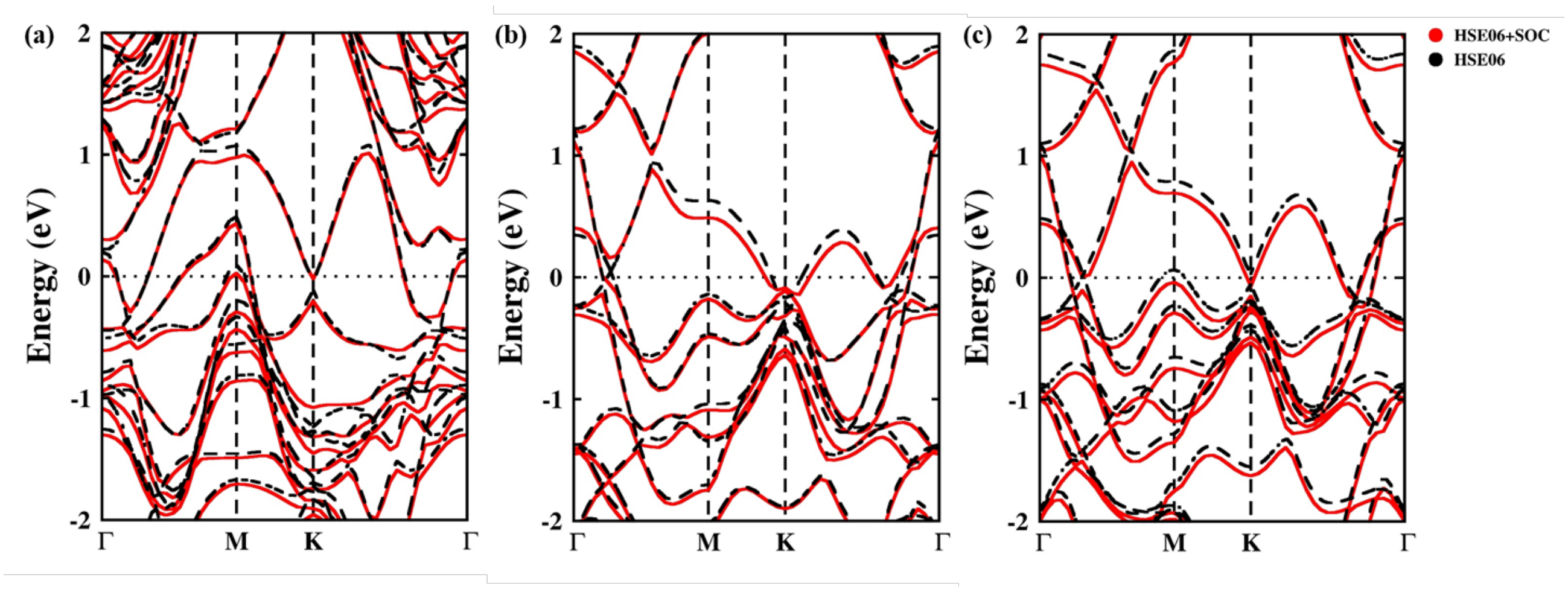

**Fig. S7:** Electronic band structures for (a) $W_2CrC_2$ (b) $W_2CrC_2O_2$ and (c) $Mo_2CrC_2O_2$ systems with (HSE06+SOC) and without (HSE06) spin orbit coupling effect. The red and dashed black bands are representing with and without spin orbit coupling effect electronic bands structure, respectively. The Fermi energy is shifted to zero and indicated by the horizontal dashed black line.

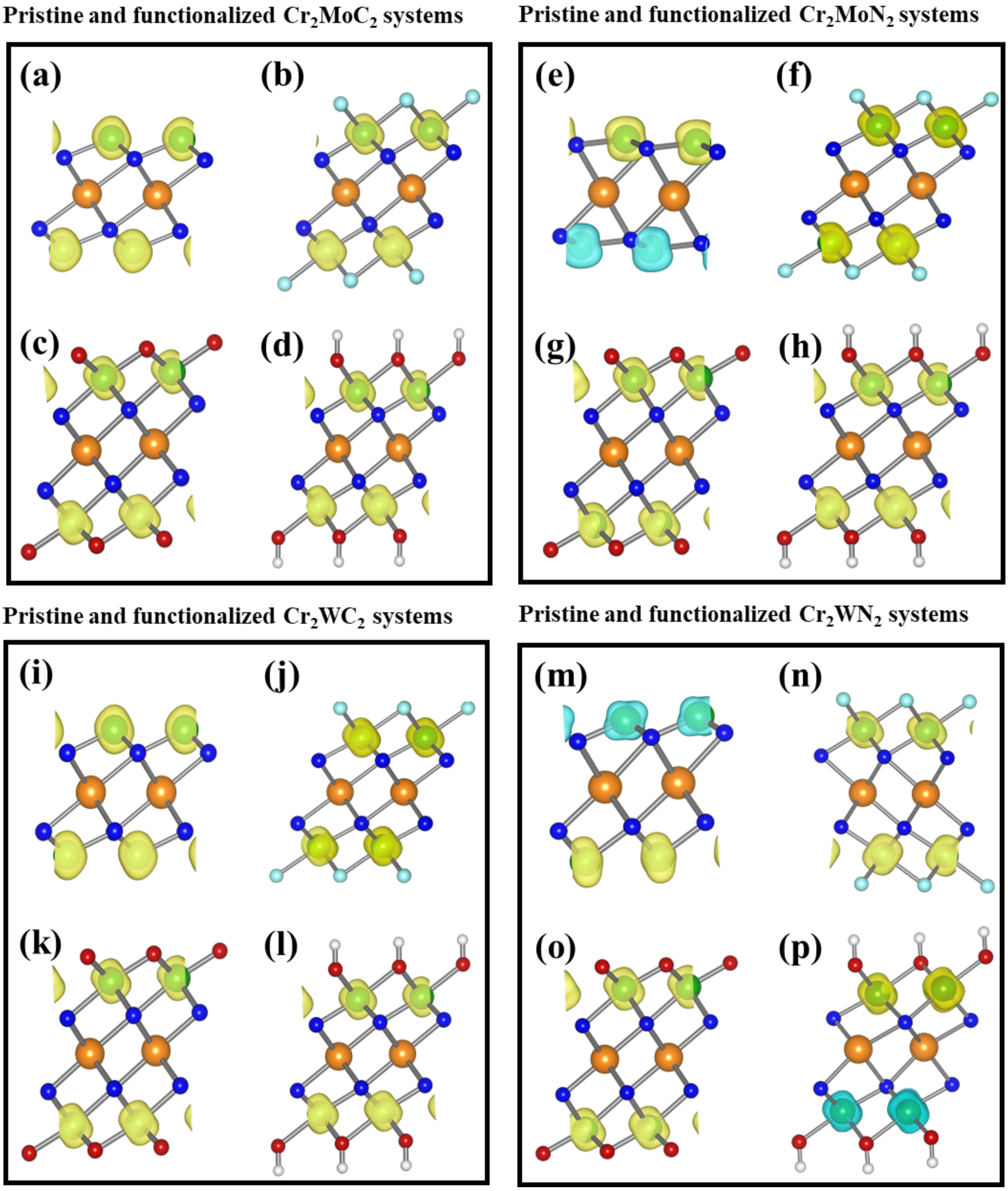

**Fig. S8:** Magnetic spin density distributions ($\Delta\rho = \rho\uparrow - \rho\downarrow$) side view for pristine $Cr_2M''X_2$ and functionalized $Cr_2M''X_2T_2$ (where M′′= Mo/W, X = C/N, and T = –F/–OH/=O) MXenes. Top panel [(a)-(h)] for (a) $Cr_2MoC_2$, (b) $Cr_2MoC_2F_2$, (c) $Cr_2MoC_2O_2$, (d) $Cr_2MoC_2(OH)_2$, (e) $Cr_2MoN_2$, (f) $Cr_2MoN_2F_2$, (g) $Cr_2MoN_2O_2$, (h) $Cr_2MoN_2(OH)_2$ MXenes. Bottom panel [(i)-(p)] for (i) $Cr_2WC_2$, (j) $Cr_2WC_2F_2$, (k) $Cr_2WC_2O_2$, (l) $Cr_2WC_2(OH)_2$, (m) $Cr_2WN_2$, (n) $Cr_2WN_2F_2$, (o) $Cr_2WN_2O_2$, and (p) $Cr_2WN_2(OH)_2$ MXenes. The yellow, and cyan isosurfaces indicate up and down spin densities, respectively.

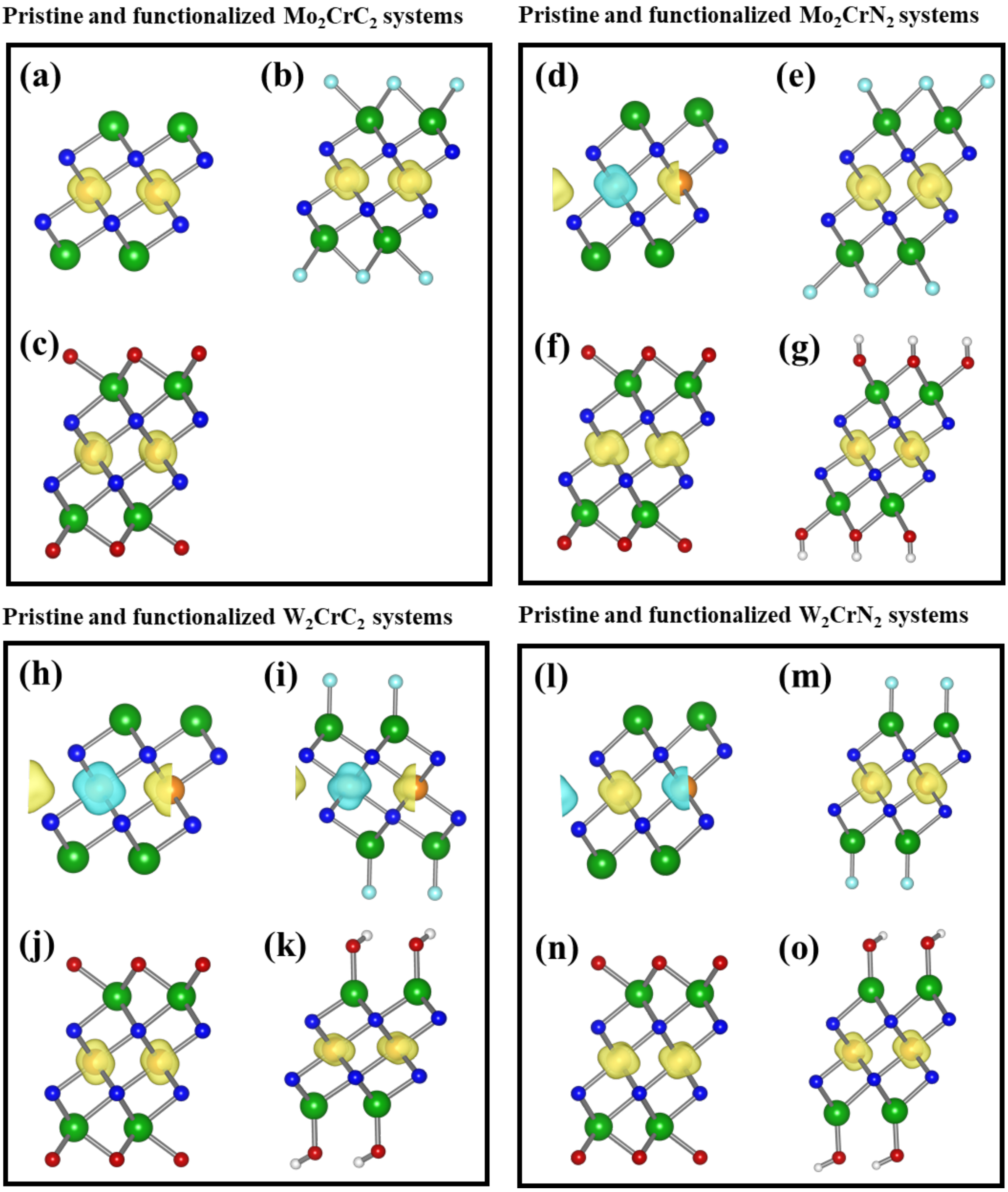

**Fig. S9:** Magnetic spin density distributions ($\Delta\rho = \rho\uparrow - \rho\downarrow$) side view for pristine $M'_2CrX_2$ and functionalized $M'_2CrX_2T_2$ (where M′= Mo/W, X = C/N, and T = –F/–OH/=O) MXenes. Top panel [(a)-(g)] for (a) $Mo_2CrC_2$, (b) $Mo_2CrC_2F_2$, (c) $Mo_2CrC_2O_2$, (d) $Mo_2CrN_2$, (e) $Mo_2CrN_2F_2$, (f) $Mo_2CrN_2O_2$, (g) $Mo_2CrN_2(OH)_2$ MXenes. Bottom panel [(h)-(o)] for (h) $W_2CrC_2$, (i) $W_2CrC_2F_2$, (j) $W_2CrC_2O_2$, (k) $W_2CrC_2(OH)_2$, (l) $W_2CrN_2$, (m) $W_2CrN_2F_2$, (n) $W_2CrN_2O_2$, and (o) $W_2CrN_2(OH)_2$ MXenes. The yellow and cyan isosurfaces indicate up and down spin densities, respectively.

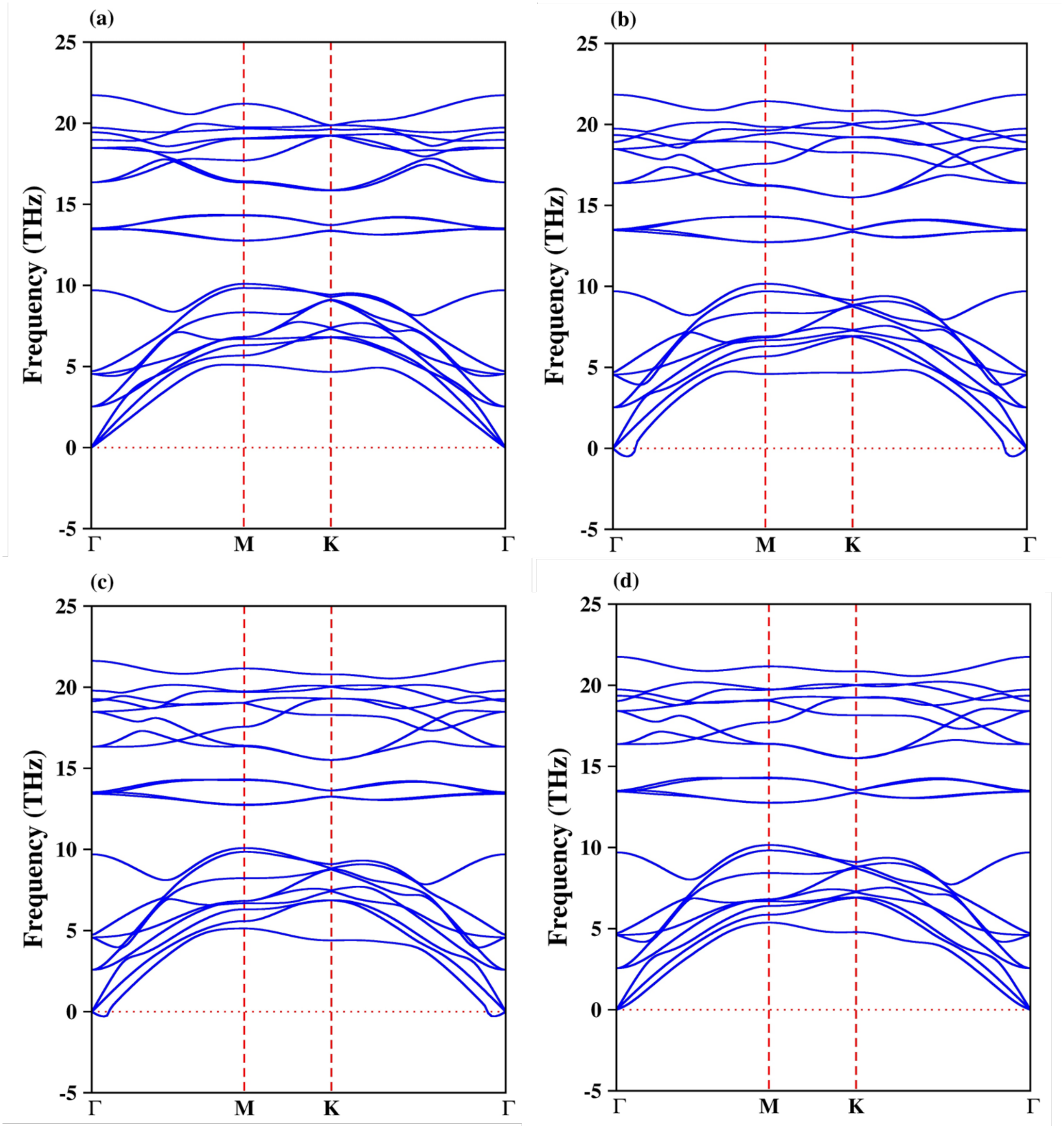

**Fig. S10:** The phonon dispersion calculated within PBE for $Mo_2CrC_2O_2$ ($H_XH_X$) MXenes (a) 2x2x1, (b) 3x3x1, (c) 4x4x1, and (d) 5x5x1 supercell with Γ-centred k-mesh of 6x6x1, 3x3x1, 2x2x1 and 2x2x1, respectively. $H_XH_X$ indicating, hollow site of C/N top for top and bottom surface of MXenes.

**Table S1:** Calculated relative energies ($E_R$ in eV) for passivation of functional groups (T = –F/–OH/=O) at different sites of MXene surfaces (Model I-IV) without (w/o) and with vdW dispersion correction, magnetic moment at ferromagnetic state (MM in $\mu_B$/unit cell), and magnetic energy difference $\Delta E = E_{AFM}-E_{FM}$ (eV) for $Cr_2MoX_2T_2$ and $Cr_2WX_2T_2$ (X = C/N) systems with four modelled structures calculated with PBE.

| **Surface termina-ting group** | **Models** | **$E_R$ w/o vdW ($E_R$ with vdW)** | **MM** | **ΔE** | **$E_R$ w/o vdW ($E_R$ with vdW)** | **MM** | **ΔE** |
|---|---|---|---|---|---|---|---|
| | | **$Cr_2MoC_2T_2$** | | | **$Cr_2WC_2T_2$** | | |
| -F | $H_XH_X$ | 0.00 (0.000) | 4.69 | 1.195 | 0.02 (0.004) | 4.72 | 1.188 |
| | $H_MH_X$ | 0.74 (0.649) | 4.72 | 0.435 | 0.00 (0.000) | 4.71 | 1.209 |
| | $H_MH_M$ | 0.03 (0.002) | 4.69 | 1.169 | 0.09 (0.032) | 4.71 | 1.119 |
| | $T_MT_M$ | 0.97 (0.895) | 2.92 | 0.061 | 1.10 (1.022) | 3.11 | 0.067 |
| -O | $H_XH_X$ | 0.54 (0.411) | 0.00 | – | 0.69 (0.567) | 0.00 | – |
| | $H_MH_X$ | 0.39 (0.332) | 1.74 | 0.264 | 0.86 (0.471) | 0.01 | 0.062 |
| | $H_MH_M$ | 0.00 (0.000) | 4.24 | 0.041 | 0.00 (0.000) | 4.27 | 0.834 |
| | $T_MT_M$ | 1.36 (1.400) | 0.00 | – | 1.54 (1.510) | 0.03 | 0.083 |
| -OH | $H_XH_X$ | 0.98 (1.008) | 4.47 | 0.222 | 1.06 (1.062) | 4.54 | 0.279 |
| | $H_MH_X$ | 0.50 (0.507) | 4.61 | 0.618 | 0.52 (0.533) | 4.66 | 0.684 |
| | $H_MH_M$ | 0.00 (0.000) | 4.58 | 1.018 | 0.00 (0.000) | 4.65 | 0.981 |
| | $T_MT_M$ | 0.29 (0.453) | 2.89 | 0.152 | 0.43 (0.578) | 3.05 | 0.144 |
| **Surface termina-ting group** | **Models** | **$E_R$ w/o vdW ($E_R$ with vdW)** | **MM** | **ΔE** | **$E_R$ w/o vdW ($E_R$ with vdW)** | **MM** | **ΔE** |
| | | **$Cr_2MoN_2T_2$** | | | **$Cr_2WN_2T_2$** | | |
| -F | $H_XH_X$ | 0.03 (0.006) | 6.32 | 1.061 | 0.23 (0.238) | 6.42 | 0.914 |
| | $H_MH_X$ | 0.00 (0.000) | 6.31 | 1.073 | 0.00 (0.000) | 5.60 | 0.867 |
| | $H_MH_M$ | 0.02 (0.025) | 6.31 | 1.050 | 0.24 (0.233) | 6.42 | 0.894 |
| | $T_MT_M$ | 1.54 (1.540) | 2.20 | -0.002 | 1.70 (1.722) | 2.44 | 0.108 |
| -O | $H_XH_X$ | 0.64 (0.649) | 0.00 | – | 0.91 (0.925) | 0.00 | – |
| | $H_MH_X$ | 0.58 (0.627) | 3.04 | 0.339 | 0.73 (0.775) | 2.50 | 0.409 |
| | $H_MH_M$ | 0.00 (0.000) | 5.39 | 1.044 | 0.00 (0.000) | 5.65 | 1.177 |
| | $T_MT_M$ | 2.90 (2.896) | 0.00 | – | 3.33 (3.426) | 0.00 | – |
| -OH | $H_XH_X$ | 0.27 (0.213) | 4.95 | 0.116 | 1.09 (0.267) | 4.26 | -0.393 |
| | $H_MH_X$ | 0.18 (0.100) | 6.16 | 0.513 | 0.68 (0.126) | 5.11 | 0.478 |
| | $H_MH_M$ | 0.00 (0.000) | 6.17 | 0.867 | 0.00 (0.000) | 5.91 | -0.099 |
| | $T_MT_M$ | 0.92 (0.895) | 2.21 | 0.030 | 0.84 (0.887) | 2.26 | -0.271 |

**Table S2:** Calculated relative energies ($E_R$ in eV) for passivation of functional groups (T = –F/–OH/=O) at different sites of MXene surfaces (Model I-IV) without (w/o) and with vdW dispersion correction, magnetic moment at ferromagnetic state (MM in $\mu_B$/unit cell), and magnetic energy difference $\Delta E = E_{AFM}$-$E_{FM}$ (eV) for $Mo_2CrX_2T_2$ and $Mo_2WX_2T_2$ (X = C/N) systems with four modeled structures calculated with PBE.

| Surface termina -ting group | Models | $E_R$ w/o vdW ($E_R$ with vdW) | MM | ΔE | $E_R$ w/o vdW ($E_R$ with vdW) | MM |
|---|---|---|---|---|---|---|
| | | $Mo_2CrC_2T_2$ | | | $Mo_2WC_2T_2$ | |
| -F | $H_XH_X$ | 0.07 (0.050) | 1.85 | 0.086 | 0.00 (0.000) | 0.00 |
| | $H_MH_X$ | 0.00 (0.000) | 1.78 | 0.178 | 0.14 (0.133) | 0.00 |
| | $H_MH_M$ | 0.19 (0.024) | 0.00 | — | 0.09 (0.095) | 0.00 |
| | $T_MT_M$ | 0.43 (0.275) | 0.95 | -0.060 | 0.04 (0.076) | 0.00 |
| -O | $H_XH_X$ | 0.00 (0.000) | 2.02 | 0.178 | 0.00 (0.000) | 0.00 |
| | $H_MH_X$ | 0.59 (0.627) | 1.88 | 0.242 | 0.65 (0.669) | 0.00 |
| | $H_MH_M$ | 1.17 (1.242) | 1.53 | 0.093 | 1.07 (1.087) | 0.00 |
| | $T_MT_M$ | 2.77 (2.908) | 2.24 | 0.220 | 2.76 (2.697) | 0.00 |
| -OH | $H_XH_X$ | 0.00 (0.000) | 0.00 | — | 0.19 (0.138) | 0.00 |
| | $H_MH_X$ | 0.18 (0.254) | 1.52 | 0.174 | 0.26 (0.244) | 0.00 |
| | $H_MH_M$ | 0.39 (0.552) | 1.29 | 0.257 | 0.35 (0.294) | 0.00 |
| | $T_MT_M$ | 0.08 (0.247) | 0.88 | 0.041 | 0.00 (0.000) | 0.00 |
| **Surface termina -ting group** | **Models** | **$E_R$ w/o vdW ($E_R$ with vdW)** | **MM** | **ΔE** | **$E_R$ w/o vdW ($E_R$ with vdW)** | **MM** |
| | | $Mo_2CrN_2T_2$ | | | $Mo_2WN_2T_2$ | |
| -F | $H_XH_X$ | 0.63 (0.579) | 1.66 | 0.512 | 0.00 (0.000) | 0.00 |
| | $H_MH_X$ | 0.00 (0.000) | 2.92 | 0.502 | 0.70 (0.102) | 0.00 |
| | $H_MH_M$ | 0.01 (0.001) | 2.92 | 0.439 | 0.70 (0.102) | 0.00 |
| | $T_MT_M$ | 0.39 (0.459) | 2.25 | 0.585 | 0.96 (0.913) | 0.00 |
| -O | $H_XH_X$ | 0.00 (0.000) | 3.54 | 0.144 | 0.00 (0.000) | 0.00 |
| | $H_MH_X$ | 0.43 (0.417) | 2.64 | -0.067 | 0.43 (0.459) | 0.00 |
| | $H_MH_M$ | 1.19 (1.131) | 2.71 | 0.320 | 1.29 (1.109) | 0.00 |
| | $T_MT_M$ | 3.13 (3.168) | 2.29 | 0.385 | 2.86 (2.975) | 0.00 |
| -OH | $H_XH_X$ | 0.93 (0.411) | 0.00 | — | 0.37 (0.439) | 0.00 |
| | $H_MH_X$ | 0.24 (0.231) | 2.62 | 0.535 | 0.11 (0.269) | 0.00 |
| | $H_MH_M$ | 0.00 (0.000) | 2.83 | 0.599 | 0.02 (0.317) | 0.00 |
| | $T_MT_M$ | 0.14 (0.172) | 2.40 | 0.548 | 0.00 (0.000) | 0.00 |

**Table S3:** Calculated relative energies ($E_R$ in eV) for passivation of functional groups (T = –F/–OH/=O) at different sites of MXene surfaces (Model I-IV) without (w/o) and with vdW dispersion correction, magnetic moment at ferromagnetic state (MM in $\mu_B$/unit cell), and magnetic energy difference $\Delta E = E_{AFM}\text{-}E_{FM}$ (eV) for $W_2CrX_2T_2$ and $W_2MoX_2T_2$ (X = C/N) systems with four modeled structures calculated with PBE.

| **Surface termina -ting group** | **Models** | **$E_R$ w/o vdW ($E_R$ with vdW)** | **MM** | **ΔE** | **$E_R$ w/o vdW ($E_R$ with vdW)** | **MM** |
|---|---|---|---|---|---|---|
| | | **$W_2CrC_2T_2$** | | | **$W_2MoC_2T_2$** | |
| -F | $H_XH_X$ | 0.25 (0.309) | 1.77 | 0.148 | 0.44 (0.482) | 0.00 |
| | $H_MH_X$ | 0.35 (0.307) | 0.01 | 0.042 | 0.41 (0.473) | 0.00 |
| | $H_MH_M$ | 0.38 (0.326) | 0.01 | 0.013 | 0.42 (0.472) | 0.00 |
| | $T_MT_M$ | 0.00 (0.000) | 1.19 | -0.051 | 0.00 (0.000) | 0.00 |
| -O | $H_XH_X$ | 0.00 (0.000) | 2.03 | 0.158 | 0.00 (0.000) | 0.00 |
| | $H_MH_X$ | 0.68 (0.707) | 1.87 | 0.209 | 0.82 (0.813) | 0.00 |
| | $H_MH_M$ | 1.30 (1.361) | 1.21 | 0.113 | 1.42 (1.379) | 0.00 |
| | $T_MT_M$ | 2.84 (2.984) | 2.29 | 0.259 | 2.88 (2.801) | 0.00 |
| -OH | $H_XH_X$ | 0.39 (0.250) | 2.28 | 0.169 | 0.57 (0.604) | 0.00 |
| | $H_MH_X$ | 0.97 (0.610) | 0.00 | – | 0.51 (0.468) | 0.00 |
| | $H_MH_M$ | 1.19 (1.048) | 1.35 | 0.162 | 0.29 (0.279) | 0.00 |
| | $T_MT_M$ | 0.00 (0.000) | 1.32 | 0.096 | 0.00 (0.000) | 0.00 |
| **Surface termina -ting group** | **Models** | **$E_R$ w/o vdW ($E_R$ with vdW)** | **MM** | **ΔE** | **$E_R$ w/o vdW ($E_R$ with vdW)** | **MM** |
| | | **$W_2CrN_2T_2$** | | | **$W_2MoN_2T_2$** | |
| -F | $H_XH_X$ | 0.73 (0.654) | 3.12 | -0.231 | 0.40 (0.466) | 0.00 |
| | $H_MH_X$ | 0.81 (0.691) | 0.03 | 0.008 | 0.36 (0.337) | 0.00 |
| | $H_MH_M$ | 0.87 (0.783) | 0.00 | – | 0.94 (0.319) | 0.00 |
| | $T_MT_M$ | 0.00 (0.000) | 2.75 | 0.492 | 0.00 (0.000) | 0.00 |
| -O | $H_XH_X$ | 0.00 (0.000) | 3.56 | 0.153 | 0.00 (0.000) | 0.00 |
| | $H_MH_X$ | 0.66 (0.641) | 2.69 | 0.366 | 0.91 (0.830) | 0.00 |
| | $H_MH_M$ | 1.60 (1.548) | 2.75 | 0.358 | 1.81 (1.720) | 0.00 |
| | $T_MT_M$ | 3.27 (3.298) | 2.39 | 0.382 | 3.58 (3.456) | 0.00 |
| -OH | $H_XH_X$ | 1.31 (1.197) | 0.01 | -1.509 | 1.09 (1.082) | 0.00 |
| | $H_MH_X$ | 1.04 (0.931) | 2.86 | 0.491 | 0.82 (0.783) | 0.00 |
| | $H_MH_M$ | 0.80 (0.755) | 3.01 | 0.506 | 1.09 (0.557) | 0.00 |
| | $T_MT_M$ | 0.00 (0.000) | 2.72 | 0.598 | 0.00 (0.000) | 0.00 |

**Data S1:** Representative input file (INCAR) for geometry optimization calculation with VASP software package by considering PBE functional. Each line giving individual statement for electronic optimization, and ionic relaxation.

```
#------------------System_Initial------------------
SYSTEM = MXene
ISTART = 0
ICHARG = 2
#------------------Convergence_Criteria----------
ENCUT = 400
LWAVE = .FALSE.
LREAL = .FALSE.
PREC = ACCURATE
ISMEAR = 0
SIGMA = 0.05
ISPIN = 2
NSW = 1000
ISIF = 3
IBRION = 2
NPAR = 4
```

**Data S2:** Representative input file (INCAR) for phonon calculation with VASP software package by considering PBE functional.

```
#------------------System_Initial-------------
SYSTEM = MXene
ISTART = 0
ICHARG = 2
#------------------Convergence_Criteria--------
EDIFF = 1E-6
ENCUT = 520
LWAVE = .FALSE.
LREAL = .FALSE.
LCHARG = .FALSE.
ADDGRID = .TRUE.
PREC = ACCURATE
ISMEAR = 1
SIGMA = 0.15
ISPIN = 2
MAGMOM = 50*1 50*1 25*4 50*1   ## NIONS vary with size of the supercell
NSW = 0
IBRION = -1
NPAR = 7
NELM = 200
ISYM = 0
NSIM = 8
```

**Data S3:** Relaxed geometries (CONTCAR) for 36 stable MXenes using VASP.

**(1) $Cr_2MoC_2_F_2$ ($H_XH_X$)**

```
1.00000000000000
  3.0194730233058147   0.0108539064169124   0.2918603538733978
 -1.5037811454865952   2.6237183715436037  -0.2928616431855917
  2.9606693337342955  -1.9535371773001933  17.3838289465346811
C   Cr   Mo   F
2    2    1    2
Direct
0.1194396958149564  0.1693358689781292  0.2976186392154802
0.9413409858524147  0.3250139136849980  0.1401196160221075
0.7333849419436294  0.5655323876281098  0.3533093964489956
0.3300509017649078 -0.0716338848329798  0.0844047675543398
0.5320693059304588  0.7486674929430002  0.2189144305194618
0.3242334808838775 -0.0204445166479224  0.4258029291521790
0.7292306858097475  0.5120386822466600  0.0117201890874381
```

**(2) $Cr_2MoC_2_O_2$ ($H_MH_M$)**

```
1.00000000000000
  2.9598016074086977   0.0001294751763667  -0.0159568213784646
 -1.4813202887517949   2.5618581313615745   0.0154869636016760
  0.0871480041214914  -0.1776521112038440  26.1958376399060739
C   Cr   Mo   O
2    2    1    2
Direct
0.1930761077387415  0.0784565747661857  0.2690541127926327
0.8567623340500105  0.4052649067069106  0.1687767496568261
0.8692821331062793  0.4106448266403501  0.3168901367484552
0.2002408560293898  0.0578956315122822  0.1209554475047660
0.5129029870734659  0.7510393465485256  0.2189423698071800
0.5338039391215369  0.7469605873168456  0.3516613478553254
0.5262116228805777  0.7290580625088988  0.0861097976348180
```

**(3) $Cr_2MoC_2_(OH)_2$ ($H_MH_M$)**

```
1.00000000000000
  3.0567902819097985  -0.0016172912086575  -0.0161164416581772
 -1.5314333418193813   2.6441524652413482   0.0114311341463493
  0.1069298802025813  -0.2395672859522881  28.6487503518898130
C   Cr   Mo   O   H
2    2    1    2    2
Direct
0.1959526194282371  0.0790692453178504  0.2654806724620090
```

0.8573890335660891 0.4055593382194673 0.1742042859270105
0.8638492741608379 0.4141155812363082 0.2977282044548590
0.1894627170855080 0.0694406808867405 0.1415117227621897
0.5268511204710323 0.7421216577247461 0.2199032586574130
0.5348224638363938 0.7518976316365765 0.3406490843250859
0.5200417688140400 0.7332939431644292 0.0984272259120632
0.5486593863019362 0.7616486845111096 0.3745288084859072
0.5020416483359261 0.7486832853027675 0.0646766850134616

**(4) $Cr_2MoN_2$_$F_2$ ($H_MH_X$)**

1.00000000000000
2.9742931160769444 -0.0041096748836750 -0.1443387239282156
-1.4919052459824909 2.5858776823476926 -0.0018441216688113
-0.9883680580331934 -1.0853584017909526 20.7008627066536093
N Cr Mo F
2 2 1 2
Direct
0.3602683433634011 0.1744728747538881 0.2870004018249818
0.8763713506470792 0.4060482336549092 0.1484034080661537
0.0780736807092199 0.5423749198971660 0.3360743545264526
0.1591438036577398 0.0395506536260177 0.0989529044482292
0.6109417309997665 0.7850498765723449 0.2174701648970881
-0.1921112791672815 -0.0823043105359659 0.3939543629172935
0.4375123527900744 0.6709677080316390 0.0410643453198023

**(5) $Cr_2MoN_2$_$O_2$ ($H_MH_M$)**

1.00000000000000
2.9086887927084666 0.0023455572990920 -0.0341579871781104
-1.4621686528419409 2.5169185336942284 0.0084300689702202
0.1069966059623223 -0.0774774935235882 23.7125557719056808
N Cr Mo O
2 2 1 2
Direct
0.2490523726755754 0.2215262719465850 0.2481146917089875
0.9051392402845710 0.5476130554191224 0.1336501167621903
0.9204870474252156 0.5580982713677733 0.3036792650796401
0.2344270585993006 0.2108998868575251 0.0781126722939640
0.5780417500977474 0.8846984173056966 0.1908996363346655
0.5909654181693168 0.8936199556614939 0.3415219828663195
0.5639971687482790 0.8752542574418064 0.0403316839542238

**(6) $Cr_2MoN_2$_$(OH)_2$ ($H_MH_M$)**

1.00000000000000
2.9922972893810069 0.0001809684054022 -0.0163524874786575
-1.4975843046677575 2.5921504762168048 0.0120798149851357
0.0906232890790099 -0.2247250512891717 26.4919591092428703
N Cr Mo O H
2 2 1 2 2
Direct
0.1962504986449595 0.0859103576723487 0.2729961278124594
0.8558914041445427 0.4101548539699346 0.1664866224350111
0.8661310072275565 0.4223305574799776 0.3116770791475098
0.1860254260836008 0.0745794941287580 0.1276745713969569
0.5267800283688978 0.7496890859976858 0.2197080434130121
0.5353298930786878 0.7575948969274028 0.3562029915250972
0.5171066578661121 0.7364573572591725 0.0830716158868955
0.5278519882461401 0.7507822230603151 0.3929372802503049
0.5277031283395033 0.7183312215044019 0.0463556161327517

**(7) $Cr_2WC_2$_$F_2$ ($H_MH_X$)**
1.00000000000000
3.0421441137850387 -0.0046168858022502 -0.1630332238016560
-1.5274809875490496 2.6346954280774040 -0.0057631910629097
-0.9570525101531792 -0.9662650488680817 17.5163704941593110
C Cr W F
2 2 1 2
Direct
0.3572837246085680 0.1708956951915646 0.2930376568991265
0.8818962444231149 0.4132139601562380 0.1421478665490490
0.0742747488508077 0.5366764569979663 0.3474381381649529
0.1655083672936730 0.0473353933392172 0.0876320569707024
0.6197573828449566 0.7915929986644337 0.2176387265249684
-0.1981738024872589 -0.0918891821717981 0.4179720535665383
0.4296533174661380 0.6683346338223778 0.0170534433246644

**(8) $Cr_2WC_2$_$O_2$ ($H_MH_M$)**
1.00000000000000
2.9788542630154247 -0.0002326615809175 -0.0153602832448907
-1.4911749813620416 2.5792682720838944 0.0142804657010741
0.0931534059280480 -0.1886431929588830 26.1868029324750147
C Cr W O
2 2 1 2
Direct
0.1907433771905257 0.0806663929031478 0.2681863860117122
0.8547400661566502 0.4066608174203419 0.1697190122768444
0.8732459920048798 0.4069882205649991 0.3160596031679176

0.2033823417971101 0.0555032252824617 0.1216917240599102
0.5075447012717894 0.7556721859666022 0.2189605332890497
0.5344963283140287 0.7461883209408413 0.3505570599609403
0.5281271732650187 0.7276407729216047 0.0872156432336289

**(9) $Cr_2WC_2_(OH)_2$ ($H_MH_M$)**

1.00000000000000
3.0581745762254977 -0.0018081793876647 -0.0166407017721858
-1.5322896280507610 2.6466603975277554 0.0117504436076580
0.1023165925024946 -0.2392492249549623 28.6852343801914493
C Cr W O H
2 2 1 2 2
Direct
0.1956346717197941 0.0792309610165515 0.2648174653437195
0.8576107010325223 0.4055012178678123 0.1744272118557983
0.8641369258458640 0.4142948004229189 0.2979983990592098
0.1893033544096145 0.0694100830263005 0.1413825845365747
0.5264855803447076 0.7424632044846903 0.2196064020766245
0.5351398519280099 0.7517153615080701 0.3408869363603369
0.5198951310037715 0.7336009020769333 0.0984858114231385
0.5483773278167322 0.7614447830676692 0.3747889985686110
0.5024864878989841 0.7481687345290500 0.0647161387759861

**(10) $Cr_2WN_2 F_2$ ($H_MH_X$)**

1.00000000000000
2.9244640698208686 0.0132248006028707 -0.2329388713731469
-1.4552478141816962 2.5365777555913551 0.2325954598582269
-1.7990278207399069 0.9198383821455168 22.6565945197397660
N Cr W F
2 2 1 2
Direct
0.3414581531758745 -0.0707542019607121 0.2818873253014664
0.8958363211123140 0.3662450420233130 0.1537013690971859
0.7178557719487050 0.5570951660225775 0.3302649723848399
0.1871251933247064 0.0715916764940173 0.1053546983659641
0.6144338665905141 0.6518033705353120 0.2159375196754988
0.0993491963592930 0.1794291039784031 0.3843807783615706
0.4741414804885934 0.7807497989070876 0.0513932788134757

**(11) $Cr_2WN_2_O_2$ ($H_MH_M$)**

1.00000000000000
2.9123786122883404 0.0001539044100996 -0.0268149798443073
-1.4659618427849215 2.5184289392100938 0.0026903109510550
0.1712535417009408 -0.0963439959125969 23.9285987267878539

N Cr W O
2 2 1 2
Direct
0.2498255483816884 0.2212114011791978 0.2466382536469763
0.9056116523003369 0.5486829513741712 0.1351849132389906
0.9218927648386019 0.5573504593431076 0.3041204387513190
0.2327936238387277 0.2109773771267351 0.0776565212515293
0.5772733542851263 0.8851810426888047 0.1909131697167369
0.5909276229922481 0.8933343545455688 0.3410073590938854
0.5637854893632765 0.8749725297424178 0.0407893933005534

**(12) $Cr_2WN_2$_$(OH)_2$ ($H_MH_M$)**
1.00000000000000
2.9899851138447882 -0.0010222027326404 -0.0161473974968597
-1.4974750023592380 2.5898373897329128 0.0113956116649227
0.1035107334998055 -0.2350091260358458 28.6972061037665149
N Cr W O H
2 2 1 2 2
Direct
0.1975966384049578 0.0811424873594501 0.2673761994983666
0.8546248325434476 0.4091533754910839 0.1714682551894309
0.8677642650481256 0.4166110330994531 0.3046687066683931
0.1851476287787165 0.0729962421145079 0.1347279537599458
0.5280126944907110 0.7428805257982466 0.2194391005517270
0.5370558366589538 0.7534438697165996 0.3453424125463576
0.5151040116332877 0.7373771122268893 0.0943941493920601
0.5401446591215375 0.7565324521272028 0.3792837430771271
0.5136194653202635 0.7356929500665619 0.0604094273165909

**(13) $Mo_2CrC_2$_$F_2$ ($H_MH_X$)**
1.00000000000000
2.9679305371783502 -0.0255671378091718 -0.1131652737068239
-1.5075916181173785 2.5566719482917835 0.1125814228908506
-0.7506326346122515 0.3052182028564713 21.6781121766917195
C Mo Cr F
2 2 1 2
Direct
0.1768808999038275 0.0940353828596949 0.2728540255880242
0.8009514822281816 0.4615453024990148 0.1622487100514595
0.8657566865825436 0.4095417025235997 0.3290649245453935
0.1041248714167482 0.1539378400005493 0.1062034547863806
0.4881762356222244 0.7784549081106288 0.2185956300098354
0.2324535517924203 0.0483072846294797 0.3976689678781675
0.6618562554540539 0.5903375353770318 0.0362842291407409

**(14) $Mo_2CrC_2_O_2$ ($H_XH_X$)**

```
   1.00000000000000
     2.9127283337619838    0.0006052630092376   -0.0054000980748417
    -1.4574103145826991    2.5219285662496489    0.0031055651716204
     0.2197827052903442   -0.2607921515705078   24.6156797055470378
   C    Mo   Cr   O
   2     2    1   2
Direct
  0.1955941961213418  0.0850198659126497  0.2681710225134262
  0.8633486838776386  0.4081304774644515  0.1695131044215530
  0.8629710788169808  0.4247554546419873  0.3220126134446884
  0.1984157856971235  0.0701609411589931  0.1156585606266663
  0.5270665580068811  0.7448229365053122  0.2188412045802054
  0.1963166041144398  0.0968427868631898  0.3713550114320490
  0.8660370913655938  0.3987774814534120  0.0663384509814063
```

**(15) $Mo_2CrC_2_(OH)_2$ ($H_XH_X$)**

```
   1.00000000000000
     2.9551836975286139   -0.0040589408615725   -0.0042961925849719
    -1.4826940963525517    2.5551634571866488    0.0017434330582069
     0.2217269243411139   -0.2674243597296734   27.7449550361706905
   C    Mo   Cr   O    H
   2     2    1   2    2
Direct
  0.1999756101796344  0.0871879325236573  0.2623437724944069
  0.8645998823423067  0.4120706520677174  0.1766023109346419
  0.8658217069604852  0.4252252622893538  0.3064548456091137
  0.1968382799889475  0.0736192212254667  0.1324882693004092
  0.5340585171908188  0.7500482856757209  0.2194416763965397
  0.1981395203093730  0.0961721849090969  0.3574454977409783
  0.8632464037577789  0.4030277791896380  0.0815675742174253
  0.1981937490700429  0.0932679504205425  0.3928990024243019
  0.8705962912006000  0.4111007186988008  0.0460469958821831
```

**(16) $Mo_2CrN_2_F_2$ ($H_MH_X$)**

```
   1.00000000000000
     2.8505191093052047    0.0035679621503344    0.1203447424322376
    -1.4242678512748910    2.4724710645780119   -0.1246864851103163
     1.3300217063632265   -0.9391822526181156   24.7697394706260141
   N    Mo   Cr   F
   2     2    1   2
Direct
  0.1177535376426385  0.1628071986220019  0.2694585538591243
```

0.8392236004440640 0.4254618933143721 0.1682623577466447
0.7575448492048810 0.5309380220234309 0.3244300664482694
0.2048893992555536 0.0556742209504283 0.1132062971591684
0.4859917711492521 0.7928740712877246 0.2187695922067266
0.3843986110800499 -0.0901782094980814 0.3857563056983606
0.5688282102235618 0.6808927623001241 0.0520267798817092

**(17) $Mo_2CrN_2$_$O_2$ ($H_XH_X$)**

1.00000000000000
2.8810515440279749 0.0061213431664169 0.0087722504226989
-1.4457528674633608 2.4911168643091153 -0.0093905761697199
0.4144506110124115 0.0108253223820867 21.0001722157750770
N Mo Cr O
2 2 1 2
Direct
0.2199509959956426 0.2731224377988160 0.2962564693536437
0.9021110607586926 0.6078140569590108 0.1735883351763276
0.8761672253809127 0.6076656200371598 0.3600683341082240
0.2457987411070736 0.2734447470884330 0.1097924820550407
0.5654493104372098 0.9375001828632701 0.2349809876654563
0.9219781624509297 0.6055646607198868 0.0526974396411672
0.2002045528695476 0.2753184505334292 0.4171759690001369

**(18) $Mo_2CrN_2$_$(OH)_2$ ($H_MH_M$)**

1.00000000000000
2.8914682416682660 -0.0041143335920487 -0.0254974326206795
-1.4508058476189867 2.5015445678114405 0.0211007965675573
-0.0284098746516607 -0.1524438719111458 26.7033710265827864
N Mo Cr O H
2 2 1 2 2
Direct
0.1985090811120470 0.0785283797804402 0.2652337527692896
0.8534240869850711 0.4108815093678852 0.1740775785149554
0.8745559328798266 0.4100196747697429 0.3164614125990372
0.1783510298409903 0.0801770888484501 0.1229433318088360
0.5248998298251039 0.7460791154495798 0.2197130227427747
0.5435055248088907 0.7479555446103932 0.3705163145749629
0.5109535267782875 0.7412196846166287 0.0688617156148214
0.5215315530786840 0.7735101961846415 0.4069004343441465
0.5333394666910989 0.7174588543722333 0.0324023850311751

**(19) $Mo_2WC_2$_$F_2$ ($H_XH_X$)**

1.00000000000000
3.0483628251964485 -0.0377511985246832 0.1303960080720092

-1.5587856491976539 2.6190171945339187 -0.1304225336367184
1.4898796552192564 -0.9718583022550447 23.6422508098756019
C Mo W F
2 2 1 2
Direct
0.1676955553441953 0.1142376587097624 0.2749255459124728
0.8913990167737432 0.3800331764610550 0.1627504046376424
0.8189485341164480 0.4679755449792394 0.3233140238614063
0.2423155222305222 0.0267524479296431 0.1143474926520801
0.5296431340725737 0.7460827942518815 0.2187774861699717
0.2446549172184048 0.0474386356583695 0.3889972092851163
0.8150933182441061 0.4459896860100451 0.0487778054813119

**(20) $Mo_2WC_2_O_2$ ($H_XH_X$)**
1.00000000000000
2.9113559213125115 0.0000751708663267 -0.0021975001001110
-1.4571688105607685 2.5200692177614070 -0.0004633291760816
0.2579294019352728 -0.2883831023801108 26.9070051631350218
C Mo W O
2 2 1 2
Direct
0.1958270690470237 0.0850968562685080 0.2710012814858642
0.8632031482794627 0.4081302304470389 0.1666746064196657
0.8626567499341374 0.4242840137213129 0.3211792268306690
0.1975397755458326 0.0698387164592681 0.1165023173852777
0.5294509833756676 0.7464037035453785 0.2188584900871673
0.1957102360435760 0.0966023739734741 0.3663904865763299
0.8653620357742923 0.3981540495850153 0.0712835592150279

**(21) $Mo_2WC_2_(OH)_2$ ($T_MT_M$)**
1.00000000000000
3.0366699624141682 -0.0039663584098953 0.0658522532957345
-1.5218501201181358 2.6377503328346994 0.0322312165165011
0.8538566694311811 0.4693526247388432 26.7039445524386423
C Mo W O H
2 2 1 2 2
Direct
0.1651310490373846 0.0215693171922354 0.2655883439482930
0.8991480663375084 0.3953609941648671 0.1681654460015915
0.8065871924168406 0.3440059648631335 0.3093765019740154
0.2635515957246758 0.0847025371857656 0.1245401264559939
0.5331309123305724 0.7101876773259772 0.2168088473575677
0.7635315233298413 0.2996659460869763 0.3823579054445906
0.3151670465565177 0.1129469208275228 0.0517916597162364

0.9767672725710596 0.6240539096165835 0.3980666808135160
0.0054852326955981 0.0515466397369384 0.0362344122881983

**(22) $Mo_2WN_2_F_2$ ($H_XH_X$)**

1.00000000000000
2.8138659315810952 0.0185266846240602 -0.0112841531624515
-1.4008804698978625 2.4088779339300883 0.2987010393422840
0.0604467227970783 3.1354726463010483 22.6238693271372568
N Mo W F
2 2 1 2
Direct
0.2797964067241776 0.2530757048762545 0.2996771998791266
0.8197102428415393 0.3187630280729052 0.1446972816270467
0.8728778026239687 0.4459564238281864 0.3589486313627938
0.2291086956956050 0.1397550788963078 0.0795860041960063
0.3985685439554411 0.4860997542596048 0.2105451057402947
0.4572707022552003 0.6118852247012992 0.4275217888868472
0.6524176039040598 -0.0270252706345627 0.0109139563078860

**(23) $Mo_2WN_2_O_2$ ($H_XH_X$)**

1.00000000000000
2.8374493803295677 -0.0049084244782584 0.0029619870437912
-1.4184915215224816 2.4600485209586371 -0.0019265175348771
0.1916599580663285 -0.2513239619473481 23.7984173138130259
N Mo W O
2 2 1 2
Direct
0.2113621340125803 0.0966649630145744 0.2824533629420866
0.8743995938772915 0.4178219234746732 0.1551086052701595
0.8654292291160102 0.4280428202967140 0.3412640228873901
0.2001282837563388 0.0702894251178346 0.0964227392723875
0.4921185429591801 0.7166768949445388 0.2188990223806637
0.1969291546794097 0.1005550197725681 0.3928629591500282
0.8693730665991903 0.3984690503790977 0.0448792920972747

**(24) $Mo_2WN_2_(OH)_2$ ($T_MT_M$)**

1.00000000000000
2.8402522642995747 -0.0047118586327211 0.0597650298177147
-1.4257273253277332 2.4661367707613429 0.0187262373284937
0.8589128051506812 0.3451442428243934 28.3510379937947086
N Mo W O H
2 2 1 2 2
Direct

0.1450182849802906 -0.0058385843952194 0.2691349930460498
0.8994739204899577 0.3790682260404413 0.1648399518285232
0.7773188507826727 0.3386782588028157 0.3193413791300900
0.2770419510874855 0.0936721445860371 0.1144269523384698
0.5318818455475574 0.7445067055175977 0.2169367571323223
0.7280277775012053 0.3233231743371460 0.3872080084443488
0.3345367551444169 0.1306105687753732 0.0467144528429272
0.0523380714638072 0.6175429067019961 0.4012300610570316
-0.0171375659973950 0.0224765066338122 0.0330973681802401

**(25) $W_2CrC_2$_$F_2$ ($T_MT_M$)**

1.00000000000000
3.0054912609749329 -0.0011020595323671 -0.0069634004256382
-1.5053116513772637 2.6013176425358218 0.0092142820670801
0.1991177016379197 -0.2103688724279944 27.4347679613102358
C W Cr F
2 2 1 2
Direct
0.1918810481454369 0.0755281827541978 0.2601366674635803
0.8592884047168344 0.4050446802654692 0.1752659225469046
0.8586811020651010 0.4098185205551184 0.3021945837805285
0.1923958267500837 0.0698036549792635 0.1332359499182647
0.5254389696257695 0.7404503451555161 0.2176914004189222
0.8584825061037308 0.4105955293139576 0.3706256677414904
0.1925520815930402 0.0645290189764719 0.0647597561303128

**(26) $W_2CrC_2$_$O_2$ ($H_XH_X$)**

1.00000000000000
2.9032949271834885 0.0013045509995570 -0.0051968028629703
-1.4520833092542278 2.5140990311113511 0.0029112944169149
0.2206896071075969 -0.2607968218407347 24.6598911680438775
C W Cr O
2 2 1 2
Direct
0.1956552324743639 0.0850502116369395 0.2683811219806857
0.8633602850967428 0.4081533199035965 0.1693000125826529
0.8628652913002229 0.4248167446156170 0.3221296992681755
0.1984579172167334 0.0700557343351441 0.1155508696298184
0.5272253973931684 0.7449479940601433 0.2188426631193577
0.1962881260104215 0.0967753498016267 0.3718990422301060
0.8658977485083462 0.3987105896469282 0.0657865591891983

**(27) $W_2CrC_2$_$(OH)_2$ ($T_MT_M$)**

1.00000000000000

3.0287747237419795 -0.0093413919622641 0.0660082725467998
-1.5224557116794908 2.6239519044734947 0.0343434496549689
0.8448172893380212 0.4094572151178283 26.2000765424494304
C W Cr O H
2 2 1 2 2
Direct
0.1699946977295806 0.0286530615622716 0.2618476857599365
0.8946208887760211 0.3994730364144432 0.1724097468963179
0.8125748526230180 0.3465377914418429 0.3051624137168906
0.2548089932645371 0.0838486976695598 0.1288611848468700
0.5342929539255563 0.7170528845433554 0.2170125967831526
0.7638791279768854 0.2901100641715222 0.3793617285190880
0.3112677830522507 0.0937141097568927 0.0547358555852345
0.9696654505174517 0.6164920365355494 0.3956866037826592
0.0173951431346978 0.0681582249045622 0.0378521081098536

**(28) $W_2CrN_2_F_2$ ($T_MT_M$)**

1.00000000000000
2.8768307190813003 0.0136449029936176 -0.0036141411801608
-1.4316424885494055 2.4956742951495583 0.0034760394578724
0.0635902176920363 -0.1063353601031048 23.5678738990266758
N W Cr F
2 2 1 2
Direct
0.1908757300714017 0.0784732908315195 0.2701757611026765
0.8601721092818463 0.4037192527087941 0.1652281936419339
0.8556622208998833 0.4145651535184298 0.3291581424057428
0.1953519102351815 0.0645468695298956 0.1062562369769129
0.5256891297527525 0.7418116094232103 0.2177005852386253
0.8574148624491108 0.4117498661119674 0.4080976307449765
0.1935539313098273 0.0609038538761742 0.0273035058891337

**(29) $W_2CrN_2_O_2$ ($H_XH_X$)**

1.00000000000000
2.8740774553179138 0.0054464761912134 0.0001336915036973
-1.4428201964595124 2.4848669509598795 -0.0011732492516855
0.3508809266806742 0.0435527679844099 21.3113486827809275
N W Cr O
2 2 1 2
Direct
0.2172155689305878 0.2754048886566766 0.2977853049787679
0.9063520317684073 0.6045492913706675 0.1720859976210102
0.8755887949353480 0.6081742905176145 0.3601870509370969

0.2473340967596118 0.2721967047514331 0.1096884837322263
0.5622902515128335 0.9396867271689205 0.2349384789222967
0.9214015061497717 0.6061757570036901 0.0524880059190346
0.2014777989434489 0.2742424965310030 0.4173866948895639

**(30) $W_2CrN_2$_$(OH)_2$ ($T_MT_M$)**

1.00000000000000
2.8930436313489496 0.0017363466442043 0.0506325148255084
-1.4456852202937285 2.4957735088377380 0.0486439888440670
0.7047810319059368 0.6622968949014131 26.8799001734008698
N W Cr O H
2 2 1 2 2
Direct
0.1566420790132269 0.0049111850290903 0.2627195913039720
0.8960051162219251 0.3966424284233239 0.1712296744560682
0.7991715234644019 0.3302382616564851 0.3145398907484545
0.2752290241502535 0.1045422499371888 0.1194249813986834
0.5289967274520928 0.7114645834558372 0.2170360646931243
0.7597807768719920 0.2935301651790825 0.3862455056762524
0.3389815416148147 0.1558308006112816 0.0477482871516448
0.9837609713279591 0.6399608427110878 0.4009397418415092
-0.0100678691166677 0.0069193899966226 0.0330461867302944

**(31) $W_2MoC_2$_$F_2$ ($T_MT_M$)**

1.00000000000000
3.0321763418085901 -0.0013204640642119 -0.0074650219053059
-1.5188465288956980 2.6239298650463097 0.0090844027680048
0.1920508380721701 -0.2132207404475435 27.1278871087198645
C W Mo F
2 2 1 2
Direct
0.1918277405639556 0.0760374975992351 0.2668764447391355
0.8596567111472325 0.4034162128804728 0.1685491403249794
0.8580429120187437 0.4121912181685952 0.3091028807548662
0.1936231021671794 0.0675265933868121 0.1262804294004217
0.5248355305494677 0.7382663352525072 0.2177096721197991
0.8567279358351541 0.4154780356782672 0.3783194194872182
0.1940060067182638 0.0628540390341050 0.0570719611735837

**(32) $W_2MoC_2$_$O_2$ ($H_XH_X$)**

1.00000000000000
2.9106671143556846 0.0035470115773323 -0.0052377359090470
-1.4538319609917871 2.5214186737743569 0.0030680093748030
0.2236707617992934 -0.2603420011871568 26.2128955145483822

C W Mo O
2 2 1 2
Direct
0.1948048419295817 0.0853288101653538 0.2729820641496680
0.8629767464137228 0.4070698600932457 0.1648211564674124
0.8627439347024685 0.4248679912896843 0.3236397436626993
0.1987799495342889 0.0701780971946423 0.1139202368523103
0.5259248467815617 0.7439611067797560 0.2187491142672472
0.1976756508842409 0.0975179426434936 0.3705102723084058
0.8668440277541276 0.3995861358338191 0.0672673802922588

**(33) $W_2MoC_2$_$(OH)_2$ ($T_MT_M$)**

1.00000000000000
3.0368419232295016 -0.0016366585992540 0.0637794675461221
-1.5201701657288083 2.6342516145413999 0.0287331675348510
0.8555320189514132 0.4236245526393939 26.9974652212564443
C W Mo O H
2 2 1 2 2
Direct
0.1658469811459577 0.0261469150185323 0.2664289875659157
0.9008250836515194 0.3996642907844943 0.1674243054952627
0.8079573202005745 0.3454962531072234 0.3088679389235288
0.2629292553753591 0.0848151860616960 0.1251194842384918
0.5302297402203459 0.7157075813959933 0.2168789763851259
0.7644133124659858 0.2944340605113052 0.3808220440691357
0.3116893713077730 0.0974720116720401 0.0532037672266610
0.9728954413044887 0.6179459885607083 0.3966656828385152
0.0117133853279945 0.0623576198880072 0.0375187372573663

**(34) $W_2MoN_2$_$F_2$ ($T_MT_M$)**

1.00000000000000
2.8217497464229324 0.0142483017272326 -0.0092187952362492
-1.4000392376723909 2.4498166502193142 0.0097625882655583
0.1183295967183361 -0.1766940523251262 27.7380489865041717
N W Mo F
2 2 1 2
Direct
0.1893462557060205 0.0791841177713603 0.2717198363470100
0.8614510419539364 0.3985488880790104 0.1637082178708225
0.8556758614925780 0.4164426113874687 0.3230563794077173
0.1953964369651301 0.0625616544946413 0.1123603471581431
0.5251783852789805 0.7437780747448193 0.2177280321159348
0.8570123295691819 0.4146694107919420 0.3901196791650272
0.1946596280341703 0.0605851747307522 0.0452174559353482

**(35) $W_2MoN_2_O_2$ ($H_XH_X$)**

```
   1.00000000000000
     2.8402603388353604    0.0077590036319767    0.0001325592452846
    -1.4237827360374689    2.4519866368678822   -0.0005718393969076
     0.3515734603853733    0.0480182119371003   23.0821546896869059
   N    W    Mo   O
   2    2    1    2
Direct
  0.2216945241899864  0.2694354159444912  0.3021675753074894
  0.9142107990066085  0.5998460585091704  0.1674970701944603
  0.8779661573245653  0.6067202293461512  0.3607311440395498
  0.2497511545075697  0.2701415382585596  0.1091629721310945
  0.5398935982411713  0.9554836813210079  0.2351710186196954
  0.9229049161215356  0.6052913652916705  0.0557193871539968
  0.2052388996085646  0.2735118673289546  0.4141108495537171
```

**(36) $W_2MoN_2_(OH)_2$ ($T_MT_M$)**

```
   1.00000000000000
     2.8382617448602687    0.0045528880553579    0.0527252800085940
    -1.4160196262203837    2.4615771943308484    0.0323371208508517
     0.7732733629289572    0.5036713053905263   27.8952668158493324
   N    W    Mo   O    H
   2    2    1    2    2
Direct
  0.1443069256670854 -0.0057138960007589  0.2705863695687175
  0.8976002103388145  0.3825232143631855  0.1631710911472679
  0.7830617044568494  0.3341793830524739  0.3215991375200041
  0.2772568481886536  0.0958296049886817  0.1123240183409769
  0.5368915642059365  0.7332345887479734  0.2169371419700873
  0.7345898129600897  0.3100002216043540  0.3903682680744129
  0.3398038907139539  0.1380400770261609  0.0435948437280966
  0.0257060163216051  0.6316182769216746  0.4048604891374287
 -0.0107170818529896  0.0243284362962548  0.0294885645130110
```

**Data S4:** Optimized structures used for lattice dynamical calculations for $Mo_2CrC_2O_2$ ($H_XH_X$), $Mo_2CrN_2O_2$ ($H_XH_X$), $W_2CrC_2O_2$($H_XH_X$), and $W_2CrN_2O_2$ ($H_XH_X$) MXenes.

**(1) $Mo_2CrC_2_O_2$ ($H_XH_X$)**

```
  1.0
    2.9100057363937517    0.0000000000000000    0.0000000000000000
   -1.4550028681968759    2.5201388928754316    0.0000000000000000
    0.0000000000000000    0.0000000000000000   24.6149922244057038
  C    Mo   Cr   O
  2    2    1    2
Direct
 0.6666666666666666  0.3333333333333333  0.0492260347595916
 0.3333333333333334  0.6666666666666667  0.9507739652404084
 0.3333333333333333  0.6666666666666666  0.1031540611578237
 0.6666666666666667  0.3333333333333334  0.8968459388421763
 0.0000000000000000  0.0000000000000000  0.0000000000000000
 0.6666666666666666  0.3333333333333333  0.1525899848742398
 0.3333333333333334  0.6666666666666667  0.8474100151257602
```

**(2) $Mo_2CrN_2_O_2$ ($H_XH_X$)**

```
  1.0
    2.8987147890419500    0.0000000000000000    0.0000000000000000
   -1.4493573945209750    2.5103606456359784    0.0000000000000000
    0.0000000000000000    0.0000000000000000   21.5421185775140707
  N    Mo   Cr   O
  2    2    1    2
Direct
 0.6666666666666666  0.3333333333333333  0.9407965787992864
 0.3333333333333334  0.6666666666666667  0.0592034212007136
 0.3333333333333333  0.6666666666666666  0.8783377362290138
 0.6666666666666667  0.3333333333333334  0.1216622637709862
 0.0000000000000000  0.0000000000000000  0.0000000000000000
 0.3333333333333333  0.6666666666666666  0.1774206269209344
 0.6666666666666667  0.3333333333333334  0.8225793730790656
```

**(3) $W_2CrC_2_O_2$ ($H_XH_X$)**

```
  1.0
    2.9121152874061176    0.0000000000000000    0.0000000000000000
   -1.4560576437030588    2.5219658176427195    0.0000000000000000
    0.0000000000000000    0.0000000000000000   24.7152055762789367
  C    W    Cr   O
  2    2    1    2
Direct
 0.3333333333333333  0.6666666666666666  0.5494769142340772
```

0.6666666666666667 0.3333333333333334 0.4505230857659228
0.6666666666666666 0.3333333333333333 0.6030573015598790
0.3333333333333334 0.6666666666666667 0.3969426984401210
0.0000000000000000 0.0000000000000000 0.5000000000000000
0.3333333333333333 0.6666666666666666 0.6526770068696877
0.6666666666666667 0.3333333333333334 0.3473229931303123

**(4) $W_2CrN_2$_$O_2$ ($H_XH_X$)**

1.0
2.8937985520210612 0.0000000000000000 0.0000000000000000
-1.4468992760105306 2.5061030594848632 0.0000000000000000
0.0000000000000000 0.0000000000000000 21.9175791711258334
N W Cr O
2 2 1 2
Direct
0.6666666666666666 0.3333333333333333 0.9386368212240868
0.3333333333333334 0.6666666666666667 0.0613631787759132
0.3333333333333333 0.6666666666666666 0.8781992870587700
0.6666666666666667 0.3333333333333334 0.1218007129412300
0.0000000000000000 0.0000000000000000 0.0000000000000000
0.3333333333333333 0.6666666666666666 0.1771968929054353
0.6666666666666667 0.3333333333333334 0.8228031070945647